\documentclass[prd,superscriptaddress,nofootinbib,showpacs,
               showkeys,preprintnumbers,twocolumn]{revtex4}
\usepackage{graphicx,dcolumn}
\begin{document}

\title{Improved Analysis of \boldmath{$J/\psi$} Decays into a Vector 
Meson and Two Pseudoscalars}

\author{Timo A. L\"ahde}
\email{talahde@pcu.helsinki.fi}
\affiliation{Helmholtz-Institut f\"ur Strahlen- und Kernphysik (HISKP),
             Bonn University, \\ Nu{\ss}allee 14-16, D-53115 Bonn,
             Germany}

\author{Ulf-G. Mei{\ss}ner}
\email{meissner@itkp.uni-bonn.de} 
\affiliation{Helmholtz-Institut f\"ur Strahlen- und Kernphysik (HISKP),
             Bonn University, \\ Nu{\ss}allee 14-16, D-53115 Bonn,
             Germany}
\affiliation{Forschungszentrum J\"ulich, Institut f\"ur 
             Kernphysik~(Th), \\ D-52425 J\"ulich, Germany}

\begin{abstract}
Recently, the BES collaboration has published an extensive partial 
wave analysis of experimental data on
$J/\psi \rightarrow \phi\pi^+\pi^-$,
$J/\psi \rightarrow \omega\pi^+\pi^-$,
$J/\psi \rightarrow \phi K^+K^-$ and
$J/\psi \rightarrow \omega K^+K^-$. These new results are analyzed 
here, with full account of detection efficiencies, in the framework of 
a chiral unitary description with coupled-channel final state 
interactions between $\pi\pi$ and $K\bar K$ pairs. The emission of a 
dimeson pair is described in terms of the strange and nonstrange 
scalar form factors of the pion and the kaon, which include the final 
state interaction and are constrained by unitarity and by matching to 
the next-to-leading-order chiral expressions. This procedure allows for 
a calculation of the $S$-wave component of the dimeson spectrum 
including the $f_0(980)$ resonance, and for an 
estimation of the low-energy constants of Chiral Perturbation Theory, 
in particular the large $N_c$ suppressed constants $L_4^r$ and $L_6^r$. 
The decays in question are also sensitive to physics associated with 
OZI violation in the $0^{++}$ channel. It is found that the $S$-wave 
contributions to $\phi\pi^+\pi^-$, $\phi K^+K^-$ and $\omega\pi^+\pi^-$ 
given by the BES partial-wave analysis may be very well fitted up to a 
dimeson center-of-mass energy of $\sim 1.2$~GeV, for a large and 
positive value of $L_4^r$ and a value of $L_6^r$ compatible with zero. 
An accurate determination of the amount of OZI violation in the $J/\psi 
\rightarrow \phi\pi^+\pi^-$ decay is achieved, and the $S$-wave 
contribution to $\omega K^+K^-$ near threshold is predicted.
\end{abstract}

\preprint{HISKP-TH-06/16}

\maketitle

\section{Introduction}

\begin{figure}[t]
\includegraphics[width=.9\columnwidth]{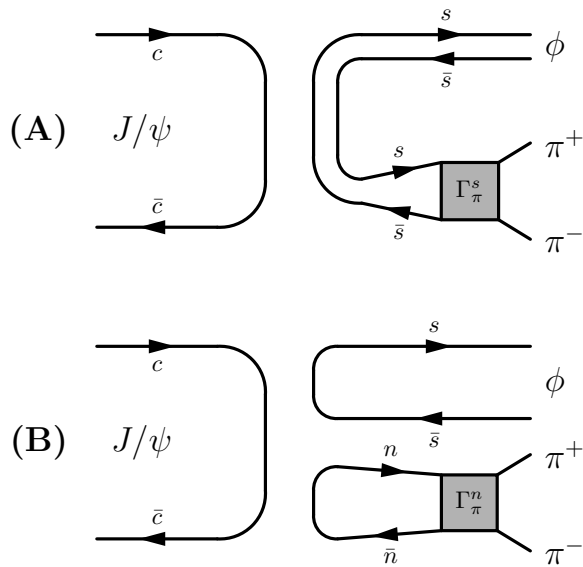}
\caption{Quark line diagrams for the decay of a $J/\psi$ into a light 
vector meson~($\phi$) and a pair of light pseudoscalar 
mesons~($\pi^+\pi^-$). The annihilation of the $c\bar c$ pair proceeds 
via at least three gluons. The shaded squares denote the 
strange~($\Gamma^s_\pi$) and nonstrange~($\Gamma^n_\pi$) scalar~FFs of 
the pion. Note that diagram~(B) is further OZI-suppressed with 
respect to diagram~(A).}
\label{Q_fig}
\end{figure}

The decays of the $J/\psi$ into a vector meson such as $\phi$ or 
$\omega$, via emission of a pair of light pseudoscalar mesons, may 
yield insight into the dynamics of the pseudo-Goldstone bosons of 
QCD~\cite{Au,Isgur1,Speth}, and in particular into the final state 
interaction (FSI) between $\pi\pi$ and $K\bar K$ pairs, which is an 
essential component in a realistic description of the scalar form 
factors~(FFs) of pions and kaons. Additionally, such decays can yield 
insight into violation of the Okubo-Zweig-Iizuka (OZI) 
rule~\cite{OZI,NC,Isg} in the scalar ($0^{++}$) sector of QCD, since 
the leading order contributions to such decays are OZI suppressed. 
Furthermore, as shown in Fig.~\ref{Q_fig}, a doubly OZI-violating 
component may contribute to these $J/\psi$ decays, which was 
demonstrated already in Refs.~\cite{UGM1,Roca} although the data from 
the DM2~\cite{DM2}, MARK-III~\cite{MK3} and BES~\cite{BES0} 
collaborations available at that time had rather low statistics. 
However, since then the BES collaboration has published far superior 
data on $J/\psi \rightarrow \phi\pi^+\pi^-$ and $J/\psi \rightarrow 
\phi K^+K^-$~\cite{BES1}, as well as for $J/\psi \rightarrow 
\omega\pi^+\pi^-$ and $J/\psi \rightarrow \omega 
K^+K^-$~\cite{BES2,BES3}. Additionally, a comprehensive partial-wave 
analysis~(PWA) has been performed for those data, which is particularly 
significant since an explicit determination of the $S$-wave $\pi\pi$ 
and $K\bar K$ event distributions is thus available. Thus, a much more 
precise analysis of the issues first touched upon in Ref.~\cite{UGM1} 
is clearly called for.

A key ingredient in such an analysis is a realistic treatment of the 
final state interaction~(FSI). It has been demonstrated in 
Ref.~\cite{OO1} that the FSI in the $\pi\pi - K\bar K$ system 
can be well described by a coupled-channel Bethe-Salpeter approach 
using the lowest order CHPT amplitudes for meson-meson 
scattering~\cite{Wein1,GL1,GL2}. In such an approach, the lowest 
resonances in the $0^{++}$ sector are of dynamical origin, i.e. they 
arise due to the strong rescattering effects in the $\pi\pi$ or $K\bar 
K$ system. Such dynamically generated states include the $\sigma$ and 
$f_0(980)$ mesons, which are prominent in the BES data on dimeson 
emission from the $J/\psi$~\cite{BES1,BES2,BES3}. It is useful, in view 
of the controversial nature of the 
$f_0(980)$~\cite{Au,Isgur1,Speth,UGM2,Tqvist,Rijken,Jaffe}, to recall 
Ref.~\cite{OO2}, which generalizes the work of Ref.~\cite{OO1}. There, 
explicit $S$-wave resonance exchanges were included together with the 
lowest order CHPT contributions in a study of the partial wave 
amplitudes for the whole scalar sector with $I=0,1/2$ and $1$. It was 
noted that the results of Ref.~\cite{OO1} could be recovered when the 
explicit tree-level resonance contributions were dropped. The 
conclusion of Ref.~\cite{OO2} was that the lowest nonet of scalar 
resonances, which includes the $\sigma,\kappa$ and $a_0(980)$ states, 
is of dynamical origin, while a preexisting octet of scalar resonances 
is present at $\sim 1.4$~GeV. It was also noted that the physical 
$f_0(980)$ state obtains a strong contribution of dynamical origin, and 
may also receive one from a preexisting singlet state.

This analysis uses the formalism introduced in Ref.~\cite{UGM1}, where 
the expressions for the scalar~FFs of the pions and kaons were obtained 
using the results of Ref.~\cite{OO1}. This allows for a description of 
the scalar~FFs which takes into account the FSI between pions and kaons 
up to $\sim 1.2$~GeV. At higher energies, a number of preexisting 
scalar resonances such as the $f_0(1500)$ have to be accounted for, as 
well as the effects of multiparticle intermediate states, most 
importantly the $4\pi$ state. These scalar~FFs may then be constrained 
by matching to the next-to-leading-order~(NLO) chiral expressions. This 
allows for a fit of the large $N_c$ suppressed Low-Energy Constants 
(LECs) $L_4^r$ and $L_6^r$ of CHPT to the dimeson spectra of the 
$J/\psi \rightarrow \phi\pi\pi$ and $J/\psi \rightarrow \phi K\bar K$ 
decays, using the Lagrangian model of Ref.~\cite{UGM1}. The amount of 
direct OZI violation present in these decays may also be accurately 
estimated. Finally, it should be noted that the present treatment of 
the FSI has been proven successful in describing, in the same spirit, a 
wide variety of processes, such as the photon fusion reactions 
$\gamma\gamma \rightarrow \pi\pi$ and $\gamma\gamma \rightarrow K\bar 
K$~\cite{phot}, the decays $\phi \rightarrow \gamma\pi\pi$ and $\phi 
\rightarrow \gamma K\bar K$~\cite{gammadec}, and more recently the 
hadronic decays $D,D_s \rightarrow \pi\pi\pi$ and $D,D_s \rightarrow 
K\pi\pi$~\cite{Ddec}.

This paper is organized in the following manner: In 
Section~\ref{theory} the description of the $J/\psi \rightarrow 
\phi\pi\pi$ decays is briefly reviewed, along with the FSI in the 
$\pi\pi - K\bar K$ system, as applied to the scalar~FFs of the 
pseudo-Goldstone bosons. Some minor corrections to the NLO CHPT 
expressions in Ref.~\cite{UGM1} are also pointed out. Section~\ref{fit} 
describes the analysis of the experimental BES data, along with a 
discussion of the fitted parameter values, with emphasis on the LECs of 
CHPT and the evidence for OZI violation. In Section~\ref{concl}, the 
results are summarized along with a concluding discussion.

\section{Theoretical Framework}
\label{theory}

The theoretical tools required for the calculation of the scalar~FFs of 
the pseudo-Goldstone bosons using CHPT and unitarity 
constraints have, as discussed in the Introduction, already been 
extensively treated in the existing literature, and therefore only 
the parts directly relevant to the present analysis are repeated here. 
For convenience, the NLO expressions for the scalar~FFs of the 
pseudo-Goldstone bosons are explicitly given here. Also, the 
expressions for the scalar~FFs of the pion in 
Ref.~\cite{UGM1} require a minor correction\footnote{The scalar~FFs 
of the pion in Ref.~\cite{UGM1} require a minor correction in the 
values at $s=0$. The authors thank J.~Bijnens and J.A.~Oller for 
their assistance in pinpointing this. The effect on the numerics of 
Ref.~\cite{UGM1} is negligible.}, such that an updated version is 
called for. A derivation of the scalar~FFs using unitarity and 
the methods of Ref.~\cite{OO1} is given in Ref.~\cite{UGM1}, and 
introductions to CHPT can e.g. be found in Ref.~\cite{ChPT}. 

\subsection{Amplitude for $\pi\pi$ and $K\bar K$ emission}

This work makes use of the SU(3) and Lorentz invariant Lagrangian of 
Ref.~\cite{UGM1} to describe the decay of a $J/\psi$ into a pair of 
pseudoscalar mesons and a light vector meson. This Lagrangian can 
be written as
\begin{eqnarray}
\mathcal{L} &=& g\,\Psi_\mu^{}\left(
\left<\mathcal{V}_8^\mu \Sigma_8^{}\right> 
+ \nu \mathcal{V}_1^\mu \Sigma_1^{}\right),
\label{Lagr}
\end{eqnarray}
where the $\mathcal{V}_8$ and $\mathcal{V}_1$ denote the lowest octet 
and singlet of vector meson resonances. Similarly, the $\Sigma_8$ and 
$\Sigma_1$ refer to the corresponding sets of scalar sources, as 
defined in Ref.~\cite{UGM1}. In the above equation, $g$ denotes a 
coupling constant, the precise value of which is not required in the 
present analysis, while the real parameter $\nu$ will be shown to play 
the role of an OZI violation parameter in the $\phi\pi^+\pi^-$ channel.
Furthermore, the angled brackets in Eq.~(\ref{Lagr}) denote the trace 
with respect to the SU(3) indices of the matrices $\mathcal{V}_8$ and 
$\Sigma_8$. Evaluation of that trace yields
\begin{eqnarray}
\mathcal{L} &=& g\,\Psi_\mu^{}\left(
V_8^\mu S_8^{} + \nu \mathcal{V}_1^\mu \Sigma_1^{} \:+\: 
\cdots\:\:\right),
\label{Lagr2}
\end{eqnarray}
where only the terms relevant for the present considerations have been 
written out. Here $V_8$ denotes the $I=0$ state of the octet of vector 
meson resonances, while $S_8$ again refers to the corresponding 
operator in the matrix of scalar sources. The $\phi$ and $\omega$ 
fields, along with the associated scalar sources $S_\phi$ and 
$S_\omega$ are then introduced according to
\begin{eqnarray}
V_8^{} = \frac{\omega}{\sqrt{3}} - \sqrt{\frac{2}{3}}\phi, \quad
S_8^{} = \frac{S_\omega}{\sqrt{3}} - \sqrt{\frac{2}{3}}S_\phi, 
\label{mixeq1} \\
\mathcal{V}_1^{} = \sqrt{\frac{2}{3}}\omega + \frac{\phi}{\sqrt{3}}, 
\quad
\Sigma_1^{} = \sqrt{\frac{2}{3}}S_\omega + \frac{S_\phi}{\sqrt{3}},
\label{mixeq2}
\end{eqnarray}
which corresponds to the ansatz of ideal mixing between $V_8^{}$ and 
$\mathcal{V}_1^{}$. The departure from this situation is reviewed, 
using different models, in Ref.~\cite{phimix}. The amount of deviation 
from ideal mixing in the $\phi -\omega$ system has been 
estimated~\cite{UGM1} to influence, in an analysis of the present kind, 
the determination of the magnitude of the OZI violation at the $\sim 
5\%$ level. This should be compared with the expected $\sim 40\%$ 
departure from unity~\cite{UGM1} of the parameter $\nu$ in 
Eq.~(\ref{Lagr2}). In view of this, the relations given in 
Eqs.~(\ref{mixeq1}) and~(\ref{mixeq2}) will be considered adequate for 
the present analysis.

The scalar sources $S_\phi$ and $S_\omega$ may, in terms of a quark 
model description, be expressed as $S_\phi = \bar ss$ and $S_\omega = 
\bar nn \equiv (\bar uu + \bar dd)/\sqrt{2}$. By means of these 
relations and Eqs.~(\ref{mixeq1}) and~(\ref{mixeq2}), the Lagrangian of 
Eq.~(\ref{Lagr2}) may be written in the form
\begin{eqnarray}
\mathcal{L} &=& \Psi_\mu\phi^\mu\:C_\phi^{}(\nu)
\left[\bar ss + \lambda_\phi^{}(\nu)\,\bar nn\right] \nonumber \\
&+& \Psi_\mu\omega^\mu\:C_\omega^{}(\nu)
\left[\bar ss + \lambda_\omega^{}(\nu)\,\bar nn\right], 
\label{Lagr3}
\end{eqnarray}
where the coupling constant $g$ is taken, as further elaborated in 
Sect.~\ref{decrat}, to be absorbed into $C_\phi$ and $C_\omega$. These 
and the $\lambda_i$ in Eq.~(\ref{Lagr3}) are given in terms of the 
parameter $\nu$ according to
\begin{eqnarray}
&& \lambda_\phi^{} = \frac{\sqrt{2}(\nu -1)}{2+\nu}, \quad
C_\phi^{} = \frac{2+\nu}{3}, \\
&& \lambda_\omega^{} = \frac{1+2\nu}{\sqrt{2}(\nu -1)}, \quad
C_\omega^{} = \frac{\sqrt{2}(\nu -1)}{3},
\end{eqnarray}
which shows that the parameters for the $\omega$ decay 
operator can be expressed in terms of those which control the $\phi$ 
decay. The explicit relations are given by
\begin{equation}
C_\omega^{} = \lambda_\phi^{} C_\phi^{}, \quad
\lambda_\omega^{} = \frac{\lambda_\phi^{} + 
\sqrt{2}}{\sqrt{2}\lambda_\phi^{}}.
\label{phirel}
\end{equation}
From now on, the dependence of the $C_i$ and $\lambda_i$ on $\nu$ will 
be suppressed. The quantities to be determined from fits to the 
experimental dimeson spectra of Refs.~\cite{BES1,BES2,BES3} are taken 
to be $C_\phi$ and $\lambda_\phi$. It is worth noting 
that the limit $\lambda_\phi = 0$ corresponds to the value $\nu = 1$, 
and in that case the dimeson spectra for $\phi$ and $\omega$ decays are
driven entirely by the strange scalar source $\bar ss$ and the 
nonstrange scalar source $\bar nn$, respectively. 

\begin{figure}[t]
\includegraphics{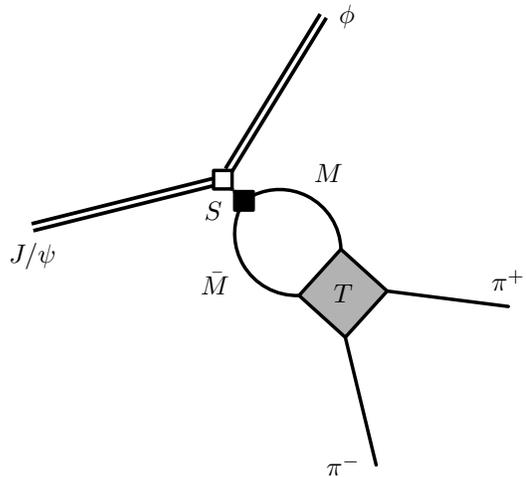}
\caption{Overview of the decay $J/\psi \rightarrow \phi\pi^+\pi^-$. The 
open square represents a vertex of the Lagrangian in Eq.~(\ref{Lagr}), 
whereas the solid square denotes the scalar~FF of the pseudoscalar 
meson pair $M\bar M$ to NLO in CHPT. The shaded square represents the 
$T$-matrix for meson-meson scattering, as calculated from the LO CHPT 
amplitudes via the Bethe-Salpeter equation. The interpolating 
field $S$ corresponds to the strange and nonstrange scalar sources 
$\bar ss$ and $\bar nn$, as described in the text.}
\label{feyn2}
\end{figure}

From the Lagrangian in Eq.~(\ref{Lagr3}), the matrix elements for
$\phi\pi^+\pi^-$ and $\phi K^+K^-$ decay of the $J/\psi$ are given by
\begin{eqnarray}
&\mathcal{M}_\phi^{\pi\pi} & = \sqrt{\frac{2}{3}}\,C_\phi^{}\,
\langle 0|(\bar ss + \lambda_\phi^{}\bar nn)|\pi\pi\rangle^*_
\mathrm{I=0}, 
\label{pipiamp} \\
&\mathcal{M}_\phi^{KK} & = \sqrt{\frac{1}{2}}\,C_\phi^{}\,
\langle 0|(\bar ss + \lambda_\phi^{}\bar nn)|K\bar K\rangle^*_
\mathrm{I=0}, 
\label{KKamp}
\end{eqnarray}
in terms of the $\pi\pi$ and $K\bar K$ states with $I=0$, which are 
related to the physical $\pi^+\pi^-$ and $K^+K^-$ states by 
the Clebsch-Gordan (CG) coefficients in front of the above expressions. 
It should be noted that in Ref.~\cite{UGM1}, the CG-coefficient for 
$\pi^+\pi^-$ decay was incorrectly written as $\sqrt{4/3}$. The full 
transition amplitudes also contain the polarization vectors 
of the $J/\psi$ and $\phi$ mesons, which have not been included 
in the above definitions. They introduce an additional, weakly energy 
dependent factor which is given explicitly in Sect.~\ref{decrat}. The 
matrix elements for $\omega\pi^+\pi^-$ and $\omega K^+K^-$, may be 
obtained by replacement of the labels in Eqs.~(\ref{pipiamp}) 
and~(\ref{KKamp}) according to $\phi \rightarrow \omega$. The matrix 
elements of the scalar sources are given in terms of the $I=0$ 
scalar~FFs which are discussed in Sect.~\ref{ffact}.

\subsection{Decay rates and dimeson event distributions}
\label{decrat}

\begin{figure*}[t]
\includegraphics[width=.95\textwidth]{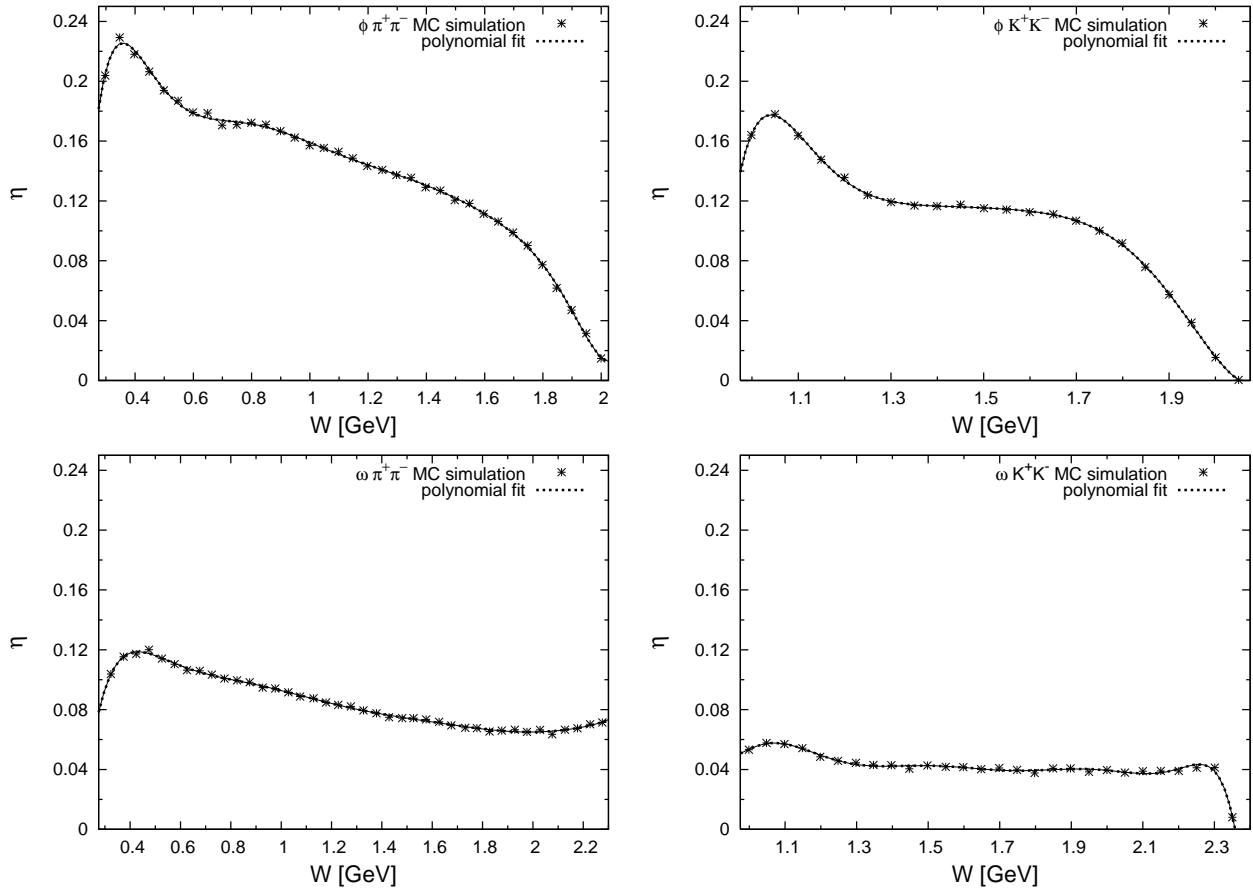}
\caption{Detection efficiency $\eta$ from BES Monte Carlo simulations 
for $J/\psi \rightarrow \phi\pi^+\pi^-$, $J/\psi \rightarrow \phi 
K^+K^-$, $J/\psi \rightarrow \omega\pi^+\pi^-$ and $J/\psi \rightarrow 
\omega K^+K^-$, along with fitted polynomial approximations for each 
case. The estimated detection efficiencies take into account 
the characteristics of the BES detector as well as the different cuts 
applied in the data analysis. The BES data published in 
Refs.~\cite{BES1,BES2,BES3} are uncorrected for efficiency, and for 
that purpose the efficiency functions $\eta$ have been included in 
Eq.~(\ref{diffev}).}
\label{eff_fig}
\end{figure*}

The differential decay rate of a $J/\psi$ into a 
vector meson and a pair of pseudoscalar mesons is, for the case of 
$\phi\pi\pi$ decay, given by
\begin{equation}
\frac{d\Gamma}{dW_{\pi\pi}} = \frac{W_{\pi\pi} |\vec p_\phi| 
|\vec p_\pi|}{4M_{J/\psi}^3\,(2\pi)^3}
\,F_{\mathrm{pol}}\,|\mathcal{M}_\phi^{\pi\pi}|^2,
\label{diffdec}
\end{equation}
where $W_{\pi\pi} = \sqrt{s_{\pi\pi}}$. The decay rates for the other 
combinations of $\phi$ and $\omega$ mesons and $\pi\pi$,$K\bar K$ final 
states can be obtained by appropriate replacement of the indices in 
Eq.~(\ref{diffdec}). The moduli of the $\phi$ and $\pi$ momenta in 
Eq.~(\ref{diffdec}) are given by
\begin{eqnarray}
|\vec p_\phi| &=& \sqrt{E_\phi^2-m_\phi^2}, \quad
E_\phi = \frac{M_{J/\psi}^2-W_{\pi\pi}^2-m_\phi^2}{2W_{\pi\pi}}, 
\quad \\
|\vec p_\pi| &=& \sqrt{E_\pi^2-m_\pi^2}, \quad
E_\pi = W_{\pi\pi}/2,
\label{mom}
\end{eqnarray}
in the rest frame of the $\pi\pi$ system. The factor $F_{\mathrm{pol}}$ 
in Eq.~(\ref{diffdec}) depends on the dipion energy $W_{\pi\pi}$ and is 
generated by properly averaging and summing over the polarizations of 
the $J/\psi$ and $\phi$ mesons, respectively. It may be expressed as
\begin{eqnarray}
F_{\mathrm{pol}} &\equiv& \frac{1}{3}\sum_{\rho,\rho'}
\varepsilon_\mu(\rho) \varepsilon^\mu(\rho')
\varepsilon_\nu^*(\rho) \varepsilon^{\nu*}(\rho') \nonumber \\
&=& \frac{2}{3}\left[1+\frac{\left(M_{J/\psi}^2 + M_\phi^2 - 
W_{\pi\pi}^2\right)^2}{8 M_{J/\psi}^2 M_\phi^2}\right].
\label{pol_eq}
\end{eqnarray}
Again, it should be noted that the corresponding expressions for 
$F_{\mathrm{pol}}$ for the other decay channels considered can be 
obtained straightforwardly from Eq.~(\ref{pol_eq}) by the substitutions
$\phi \rightarrow \omega$ and $\pi\pi \rightarrow K\bar K$. 

The results for $J/\psi$ decay into a vector meson and a dimeson pair 
published by the BES collaboration are given in terms of event 
distributions as a function of the dimeson center-of-mass energy $W$. 
The relation between the differential event distribution and the decay 
rate given by Eq.~(\ref{diffdec}) is defined to be
\begin{eqnarray}
\frac{dN}{dW_{\pi\pi}} &\sim& 
\eta(W_{\pi\pi})\,\frac{d\Gamma}{dW_{\pi\pi}},
\label{diffev}
\end{eqnarray}
where the function $\eta(W)$ represents the detection efficiency, shown 
in Fig.~\ref{eff_fig}, which also takes into account the effects of the 
various cuts imposed on the data in order to reduce the 
background to an acceptable level. It should be noted that the 
detection efficiencies $\eta$ cannot be neglected, since it is evident 
from Fig.~\ref{eff_fig} that a sizeable difference exists between the 
efficiencies for all four decays considered. Furthermore, the detection 
efficiencies exhibit significant nonlinear behavior.

The overall constant of 
proportionality in Eq.~(\ref{diffev}) is not relevant to the present 
analysis, since it may be absorbed in the definitions of $C_\phi$ and 
$C_\omega$, along with the coupling constant $g$ of Eq.~(\ref{Lagr2}) 
and a factor $\sqrt{2}B_0$ from the scalar FFs in Eq.~(\ref{FFdef}). 
Thus, the quantities with $\sqrt{2}\,gB_0$ and the proportionality 
factor absorbed are denoted by $\tilde C_\phi$ and $\tilde C_\omega$. While 
$C_\phi$ and $gB_0$ are dimensionless, $\tilde C_\phi$ has dimension 
[$E^{-\frac{1}{2}}$]. 

\subsection{Scalar Form Factors from CHPT}
\label{ffact}

The nonstrange and strange scalar~FFs of the pseudo-Goldstone 
bosons of CHPT are defined in terms of the $S$-wave states with $I=0$, 
\begin{eqnarray}
&|\pi\pi\rangle_{\mathrm{I=0}}^{} & = \frac{1}{\sqrt{3}}
\left|\pi^+\pi^-\right\rangle + \frac{1}{\sqrt{6}}
\left|\pi^0\pi^0\right\rangle, \\
&|K\bar K\rangle_{\mathrm{I=0}}^{} & = 
\frac{1}{\sqrt{2}}\left|K^+K^-\right\rangle + 
\frac{1}{\sqrt{2}}\left|K^0\bar K^0\right\rangle, \\
&|\eta\eta\rangle_{\mathrm{I=0}}^{} & = \frac{1}{\sqrt{2}}
\left|\eta\eta\right\rangle,
\end{eqnarray}
where $|\pi^+\pi^-\rangle$ denotes the symmetrized combination of 
$|\pi^+\pi^-\rangle$ and $|\pi^-\pi^+\rangle$. Following the 
conventions of Refs.~\cite{Pel1,Pel2,Pel3}, an extra factor of 
$1/\sqrt{2}$ has been included for the $I=0$ states composed of members 
of the same isospin multiplet. This takes conveniently into account the 
fact that the pions behave as identical particles in the isospin basis. 
In terms of the above states, the scalar~FFs for the $\pi$,$K$ and 
$\eta$ mesons are defined as
\begin{eqnarray}
\sqrt{2}B_0\,\Gamma^n_1(s) &=& \langle 0|\bar nn|\pi\pi 
\rangle_{\mathrm{I=0}}^{}, 
\label{FFdef} \\
\sqrt{2}B_0\,\Gamma^n_2(s) &=& \langle 0|\bar nn|K\bar K 
\rangle_{\mathrm{I=0}}^{}, 
\nonumber \\
\sqrt{2}B_0\,\Gamma^n_3(s) &=& \langle 0|\bar nn|\eta\eta 
\rangle_{\mathrm{I=0}}^{},
\nonumber
\end{eqnarray}
where the notation ($\pi$ = 1, $K$ = 2, $\eta$ = 3) has been introduced 
for simplicity. The expressions for the strange scalar~FFs may 
be obtained by the substitutions $\Gamma^n_i \rightarrow 
\Gamma^s_i$ and $\bar nn \rightarrow \bar ss$. As discussed above, the 
expressions given in Ref.~\cite{UGM1} are updated here with minor 
corrections to $\Gamma^n_1$ and $\Gamma^s_1$. With these definitions, 
the scalar~FFs may be expressed in terms of the meson loop 
function $J^r_{ii}$~\cite{GL1} and the tadpole factor $\mu_i$, given in 
Eqs.~(\ref{m_loop}) and~(\ref{tadp}) respectively. The expressions so 
obtained up to NLO in CHPT are, in agreement with 
Refs.~\cite{GL1,HB1a,HB1b},

\begin{widetext}
\begin{eqnarray}
&\Gamma_1^n(s)\:\: = & \left. \sqrt{\frac{3}{2}} \right[ 1 
+ \mu_\pi - \frac{\mu_\eta}{3}
+ \frac{16 m_\pi^2}{f^2}\left(2L_8^r-L_5^r\right)
+ 8\left(2L_6^r-L_4^r\right)\frac{2m_K^2 + 3m_\pi^2}{f^2}
+ \frac{8s}{f^2} L_4^r + \frac{4s}{f^2} L_5^r 
\nonumber \\ && + \left.
\left(\frac{2s - m_\pi^2}{2f^2}\right) J^r_{\pi\pi}(s)
+ \frac{s}{4f^2} J^r_{KK}(s)
+ \frac{m_\pi^2}{18f^2} J^r_{\eta\eta}(s)
\right], \\
&\Gamma_1^s(s)\:\: = & \frac{\sqrt{3}}{\:2} \left[  
\frac{16 m_\pi^2}{f^2}\left(2L_6^r-L_4^r\right) + \frac{8s}{f^2} L_4^r
+ \frac{s}{2f^2} J^r_{KK}(s)
+ \frac{2}{9}\frac{m_\pi^2}{f^2} J^r_{\eta\eta}(s)
\right],
\end{eqnarray}
for the pion, and
\begin{eqnarray}
&\Gamma_2^n(s)\:\: = & \left. \frac{1}{\sqrt{2}} \right[ 1 
+ \left.\frac{8 L_4^r}{f^2}
\left(2s - m_\pi^2 - 6 m_K^2\right)
+ \frac{4 L_5^r}{f^2} 
\left(s - 4 m_K^2\right)
+ \frac{16 L_6^r}{f^2} \left(6 m_K^2 + m_\pi^2\right)
+ \frac{32 L_8^r}{f^2}\,m_K^2
+ \frac{2}{3} \mu_\eta
\right. \nonumber \\ && + \left.
\left(\frac{9s - 8 m_K^2}{36f^2}\right) J^r_{\eta\eta}(s)
+ \frac{3s}{4f^2} J^r_{KK}(s)
+ \frac{3s}{4f^2} J^r_{\pi\pi}(s)
\right], \\
&\Gamma_2^s(s)\:\: = & 1 
+ \frac{8 L_4^r}{f^2} \left(s - m_\pi^2 - 4 m_K^2\right)
+ \frac{4 L_5^r}{f^2} \left(s - 4 m_K^2\right)
+ \frac{16 L_6^r}{f^2} \left(4 m_K^2 + m_\pi^2\right)
+ \frac{32 L_8^r}{f^2}\,m_K^2
+ \frac{2}{3} \mu_\eta
\nonumber \\ && +
\left(\frac{9s - 8 m_K^2}{18f^2}\right) J^r_{\eta\eta}(s)
+ \frac{3s}{4f^2} J^r_{KK}(s), 
\end{eqnarray}
for the kaon. Finally, for the $\eta$ one finds
\begin{eqnarray}
&\Gamma_3^n(s)\:\: = & \left. \frac{1}{2\sqrt{3}} \right[ 1 
+ \left.\frac{24 L_4^r}{f^2}
\left(s + \frac{1}{3} m_\pi^2 - \frac{10}{3} m_K^2\right)
+ \frac{4 L_5^r}{f^2} 
\left(s + \frac{4}{3} m_\pi^2 - \frac{16}{3} m_K^2\right)
+ \frac{16 L_6^r}{f^2} \left(10 m_K^2 - m_\pi^2\right)
\right. \nonumber \\ && + \left.
\frac{128 L_7^r}{f^2} \left(m_\pi^2 - m_K^2\right)
+ \frac{32 L_8^r}{f^2}\,m_\pi^2
- \frac{\mu_\eta}{3} + 4\mu_K - 3\mu_\pi
+ \left(\frac{16 m_K^2 - 7 m_\pi^2}{18f^2}\right) J^r_{\eta\eta}(s)
\right. \nonumber \\ && + \left.
\left(\frac{9s - 8 m_K^2}{4f^2}\right) J^r_{KK}(s)
+ \frac{3}{2} \frac{m_\pi^2}{f^2} J^r_{\pi\pi}(s)
\right], \\
&\Gamma_3^s(s)\:\: = & \left. \frac{2}{3} \right[ 1 
+ \left.\frac{6 L_4^r}{f^2}
\left(s - \frac{2}{3} m_\pi^2 - \frac{16}{3} m_K^2\right)
+ \frac{4 L_5^r}{f^2} 
\left(s + \frac{4}{3} m_\pi^2 - \frac{16}{3} m_K^2\right)
+ \frac{8 L_6^r}{f^2} \left(8 m_K^2 + m_\pi^2\right)
\right. \nonumber \\ && + \left.
\frac{64 L_7^r}{f^2} \left(m_K^2 - m_\pi^2\right)
+ \frac{32 L_8^r}{f^2} \left(2 m_K^2 - m_\pi^2\right)
- \frac{4}{3}\mu_\eta + 2\mu_K
+ \left(\frac{16 m_K^2 - 7 m_\pi^2}{18f^2}\right) J^r_{\eta\eta}(s)
\right. \nonumber \\ && + \left.
\left(\frac{9s - 8 m_K^2}{8f^2}\right) J^r_{KK}(s)
\right].
\end{eqnarray}
\end{widetext}

\subsection{Matching of FSI to NLO CHPT}

The constraints imposed by unitarity on the pion and kaon scalar~FFs, 
the inclusion of the FSI via resummation in terms of 
the Bethe-Salpeter~(BS) equation, the channel coupling between the 
$\pi\pi$ and $K\bar K$ systems, and the matching of the scalar~FFs to 
the NLO CHPT expressions have all been elaborated in great 
detail in Refs.~\cite{UGM1,OO1}, and will thus be only briefly touched 
upon here. Within that framework, consideration of the unitarity 
constraints yields a scalar~FF in terms of the algebraic 
coupled-channel equation
\begin{eqnarray}
\Gamma(s) &=& [I+K(s)\,g(s)]^{-1} R(s) \label{G_eq} \\
&=& [I-K(s)\,g(s)]\:R(s) \:+\: \mathcal{O}(p^6), \nonumber
\end{eqnarray}
where in the second line, the equation has been expanded up to NLO, 
the NNLO contribution defined as being of $\mathcal{O}(p^6)$ in the 
chiral expansion. This expansion is instructive since it allows for the 
$J^r_{ii}$ integrals from the NLO scalar~FF expressions to be absorbed 
into the above equation. Here $K(s)$ denotes the kernel of $S$-wave 
projected $I=0$ meson-meson scattering amplitudes from the 
leading order chiral Lagrangian. Using the notation defined in 
Sect.~\ref{ffact}, they are given by
\begin{eqnarray}
&& K_{11} = \frac{2s - m_\pi^2}{2f^2}, \quad
   K_{12} = K_{21} = \frac{\sqrt{3}s}{4f^2}, \label{iker} \\
&& K_{22} = \frac{3s}{4f^2}, \nonumber 
\end{eqnarray}
where the constant $f$ is taken to equal the pion decay constant 
$f_\pi$, with the convention $f_\pi = 0.0924$~GeV. The components given 
above are sufficient for the two-channel 
formalism of Ref.~\cite{UGM1} used in this paper, where only the FSI in 
the $\pi\pi$ and $K\bar K$ channels is considered. The chiral 
logarithms associated with the $\eta\eta$ channel can thus not be 
reproduced by Eq.~(\ref{G_eq}) and are therefore removed from the 
chiral expressions given in Sect.~\ref{ffact}, while the contribution 
of that channel to the values of the form factors at $s=0$ is retained. 
For completeness, it should be noted that if the $\eta\eta$ channel is 
also included, then the matrix $K(s)$ should be augmented by the 
elements
\begin{eqnarray}
&& K_{13} = K_{31} = \frac{m_\pi^2}{2\sqrt{3}f^2}, \quad
   K_{33} = \frac{16 m_K^2 - 7 m_\pi^2}{18f^2}, \nonumber \\
&& K_{23} = K_{32} = \frac{9s - 8m_K^2}{12f^2}.
\end{eqnarray}
The elements of the diagonal matrix $g_i(s)$ in Eq.~(\ref{G_eq}) are 
given by the cutoff-regularized loop integral
\begin{eqnarray}
g_i(s) &=& \frac{1}{16\pi^2} \left\{\sigma_i(s)\log\left(
\frac{\sigma_i(s) \sqrt{1 + \frac{m_i^2}{q_\mathrm{max}^2}} + 1}
     {\sigma_i(s) \sqrt{1 + \frac{m_i^2}{q_\mathrm{max}^2}} - 1}
\right)\right.\nonumber \\ 
&-& \left.2\log\left[\frac{q_\mathrm{max}^{}}{m_i}
\left(1 + \sqrt{1 + \frac{m_i^2}{q_\mathrm{max}^2}}\right)
\right]\right\}, \label{mloop} 
\end{eqnarray}
where
\begin{equation}
\sigma_i(s) = \sqrt{1-\frac{4m_i^2}{s}},
\end{equation}
and $q_\mathrm{max}$ denotes a three-momentum cutoff, which has to be 
treated as an {\it a priori} unknown model parameter, which is however 
expected to be of the order of $\sim 1$~GeV. Since the above 
expressions are calculated in a cutoff-regularization scheme, it is 
useful to note that within the modified $\overline{MS}$ subtraction 
scheme commonly employed in CHPT, the meson loop function $g_i(s)$ is 
given by 
\begin{eqnarray}
J_{ii}^r(s) \!&\equiv&\! \frac{1}{16\pi^2}\left[
1 - \log\left(\frac{m_i^2}{\mu^2}\right) - \sigma_i(s)\log\left(
\frac{\sigma_i(s)+1}{\sigma_i(s)-1}\right)\right] \nonumber \\
&=& -g_i(s),
\label{m_loop} 
\end{eqnarray}
for which it has been shown in App.~2 of Ref.~\cite{Pel1} that an 
optimal matching between the two renormalization schemes requires that 
\begin{equation}
\mu = \frac{2q_\mathrm{max}}{\sqrt{e}},
\label{mu_match}
\end{equation}
in which case the differences between the two forms are of 
$\mathcal{O}(m_i^2/q_\mathrm{max}^2)$. Furthermore, the expressions for 
the logarithms $\mu_i$ generated by the chiral tadpoles in the NLO 
scalar~FFs are given by
\begin{equation}
\mu_i = \frac{m_i^2}{32\pi^2 f^2} \log\left(\frac{m_i^2}{\mu^2}\right).
\label{tadp}
\end{equation}

As demonstrated in Ref.~\cite{UGM1}, the quantity $R(s)$ in 
Eq.~(\ref{G_eq}) is a vector of functions free of any cuts or 
singularities, since the right-hand or unitarity cut has been removed 
by construction. The information provided by CHPT can then be built 
into the formalism by fixing $R(s)$ to the NLO CHPT expressions for the 
scalar~FFs. Consideration of Eq.~(\ref{G_eq}) yields the 
defining relations
\begin{equation}
\Gamma_i^n(s) = R_i^n(s) - \sum_{j=1}^2 K_{ij}(s) g_j(s) R_j^n(s) + 
\mathcal{O}(p^6),
\label{R_rel}
\end{equation}
where it is understood that only contributions up to $\mathcal{O}(p^4)$ 
are to be retained in the product $KgR$. The analogous expressions for 
the vectors $R_i^s(s)$ associated with the strange scalar~FFs 
$\Gamma_i^s(s)$ can be obtained from the above relations by the 
substitutions $\Gamma_i^n \rightarrow \Gamma_i^s$ and $R_i^n 
\rightarrow R_i^s$. The above procedure is equivalent to the intuitive 
result obtained by dropping, in the expressions for the $\Gamma_i$, all 
occurrences of the loop integrals $J_{\pi\pi}^r$ and $J_{KK}^r$, and 
keeping only the parts of the $J_{\eta\eta}^r$ which do not depend on 
$s$. Nevertheless, the explicit 
evaluation of Eq.~(\ref{R_rel}) provides a useful check on the 
consistency of the normalization used for the NLO scalar~FFs 
and the LO meson-meson interaction kernel $K$. It should also be noted 
that the expressions for $R_1^n$ and $R_1^s$ correspond to the 
corrected scalar FFs, as explained in the beginning of 
Sect.~\ref{theory}. The expressions for the $R^i$ so obtained are

\begin{widetext}
\begin{eqnarray}
&R_1^n(s)\:\: = & \sqrt{\frac{3}{2}} \left\{ 1 + \mu_\pi - 
\frac{\mu_\eta}{3}
+ \frac{16 m_\pi^2}{f^2}\left(2L_8^r-L_5^r\right)
+ 8\left(2L_6^r-L_4^r\right)\frac{2m_K^2 + 3m_\pi^2}{f^2}
+ \frac{8s}{f^2} L_4^r + \frac{4s}{f^2} L_5^r \right.\nonumber \\
&& - \left.\frac{m_\pi^2}{288\pi^2 f^2} \left[1 + 
\log\left(\frac{m_\eta^2}{\mu^2}\right)\right]
\right\}, \\
&R_1^s(s)\:\: = & \frac{\sqrt{3}}{\:2} \left\{  
\frac{16 m_\pi^2}{f^2}\left(2L_6^r-L_4^r\right)
+ \frac{8s}{f^2} L_4^r - \frac{m_\pi^2}{72\pi^2 f^2}
\left[1 + \log\left(\frac{m_\eta^2}{\mu^2}\right)\right]
\right\},
\end{eqnarray}
for the pion, and for the kaon one finds
\begin{eqnarray}
&R_2^n(s)\:\: = & \frac{1}{\sqrt{2}} \left\{
1 + \frac{8 L_4^r}{f^2} \left(2s - 6m_K^2 - m_\pi^2\right)
+ \frac{4 L_5^r}{f^2} \left(s - 4m_K^2\right)
+ \frac{16 L_6^r}{f^2} \left(6m_K^2 + m_\pi^2\right)
+ \frac{32 L_8^r}{f^2} m_K^2 + \frac{2}{3} \mu_\eta \right.\nonumber \\
&& + \left.\frac{m_K^2}{72\pi^2 f^2}
\left[1 + \log\left(\frac{m_\eta^2}{\mu^2}\right)\right]
\right\}, \\
&R_2^s(s)\:\: = & 1 
+ \frac{8 L_4^r}{f^2} \left(s - 4m_K^2 - m_\pi^2\right)
+ \frac{4 L_5^r}{f^2} \left(s - 4m_K^2\right)
+ \frac{16 L_6^r}{f^2} \left(4m_K^2 + m_\pi^2\right)
+ \frac{32 L_8^r}{f^2} m_K^2 + \frac{2}{3} \mu_\eta \nonumber \\
&& + \frac{m_K^2}{36\pi^2 f^2} 
\left[1 + \log\left(\frac{m_\eta^2}{\mu^2}\right)\right].
\end{eqnarray}

\end{widetext}

The expressions for the $R_i^n$ and $R_i^s$ are valid when only the 
$\pi\pi$ and $K\bar K$ channels are considered in the FSI. On the other 
hand, if the full three-channel interaction kernel of Eq.~(\ref{iker}) 
is used, then the above equations should be modified such that the 
terms in square brackets are dropped. With respect to the 
omission of the $\eta\eta$ channel, it was noted in Ref.~\cite{OO2} 
that reproduction of the data on the inelastic $\pi\pi \rightarrow 
K\bar K$ cross section requires the addition of a preexisting 
contribution to the $f_0(980)$ if the $\eta\eta$ channel is included. 
On the other hand, no such contribution was found to be necessary if 
the $\eta\eta$ channel is dropped. Furthermore, the effect of this 
channel is known~\cite{OO2} to be very small for energies less than 
$\sim 0.8$~GeV. 

It should be stressed here, with respect to the above mentioned issues, 
that the main concern in the present analysis is the use of a meson-meson 
interaction kernel which is known to give a realistic description of 
the $\pi\pi$ phase shift close to the $K\bar K$ threshold. Since none 
of the adjustable model parameters have any influence on the behavior 
of the $\pi\pi$ phase shift, any model which has the proper chiral 
behavior for low energies and faithfully reproduces the $f_0(980)$ 
should therefore give similar results. In view of this, the inclusion 
or omission of the $\eta\eta$ channel, or the question of a preexisting 
contribution to the $f_0(980)$ resonance, are issues of secondary 
importance. Nevertheless, the uncertainties introduced by these issues 
into the determination of the $L_i^r$ are discussed in Sect.~\ref{fit}. 
Finally, it should be noted that in order to minimize the dependence on 
$m_\eta$, the Gell-Mann - Okubo~(GO) relation has been applied 
throughout in the polynomial terms of the $\Gamma_i$ and the $R_i$.

\section{Fits to BES Data}
\label{fit}

The event distributions given by Eq.~(\ref{diffev}) can be 
simultaneously fitted to the dimeson spectra in the $\phi\pi^+\pi^-$, 
$\omega\pi^+\pi^-$ and $\phi K^+K^-$ channels. The parameters to be 
determined via a $\chi^2$ fit are the LECs $L_4^r,L_5^r,L_6^r$ and 
$L_8^r$ which influence the scalar~FFs, as well as the model parameters 
$\tilde C_\phi$ and $\lambda_\phi$. Due to the accuracy of the BES 
data~\cite{BES1,BES2,BES3}, all of the model parameters can be well 
constrained, which is especially true for $L_4^r$ and $\lambda_\phi$, 
while the sensitivity of the fit to $L_5^r$ and $L_8^r$ turns out to be 
somewhat less. Once all the model parameters are determined by a fit to 
the three decay channels mentioned above, the event distribution in the 
remaining $\omega K^+K^-$ channel in essentially fixed. This channel is 
thus not included in the fit and is treated instead as a prediction. To 
a large extent, this also turns out to be true for the shape of the 
fitted $\omega \pi^+\pi^-$ distribution.

In spite of the above mentioned positive issues, the fitting of the 
predicted event distribution $dN/dW$ to the $S$-wave contribution from 
the PWA of the BES collaboration is complicated by several 
issues: Firstly, the detection efficiencies, determined by BES via 
Monte Carlo simulation and shown in Fig.~\ref{eff_fig}, are different 
for each decay channel, and furthermore vary appreciably over the range 
of dimeson energies considered. Secondly, the $S$-wave contribution in 
the BES PWA cannot be considered as strict experimental 
information, since it is inevitably biased to some extent by the 
parameterizations chosen there for the $\sigma$ pole. Thirdly, an 
unambiguous fit requires a highly precise description of the $f_0(980)$ 
resonance generated by the FSI, which to a large extent cannot be 
adjusted in the present model. All of these issues, as well as the 
fitted parameter values and the associated error analysis are 
elaborated in detail below.

\subsection{Definitions}
\label{fit_def}

\begin{table*}[t]
\begin{center}
\caption{Parameter values obtained from a simultaneous fit to the BES 
results~\cite{BES1,BES2,BES3} for $J/\psi \rightarrow \phi\pi^+\pi^-$, 
$J/\psi \rightarrow \phi K^+K^-$ and $J/\psi \rightarrow 
\omega\pi^+\pi^-$. All values of the $L_i^r$ are quoted at the scale 
$\mu = m_\rho$. The meson masses and other input parameters are 
given in Sect.~\ref{fit_def}, and the uncertainties of the 
parameters correspond to the error analysis described in 
Sect.~\ref{fit_err}. Fixed parameters 
not determined by the fits are given in parentheses. The fits of type~I 
optimize agreement for $J/\psi \rightarrow \phi K^+K^-$, whereas 
type~II fits do so for $J/\psi \rightarrow \phi\pi^+\pi^-$. Fits~III 
and~IV are similar to ``Fit~I" in the weighting of the input data. Note 
that ``Fit~I" represents the main result of this study, as elaborated 
in Sect.~\ref{fit}. The dimeson spectra which correspond to 
the main fits (I and II) are shown in Fig.~\ref{fit_fig}, and the 
corresponding sets of scalar~FFs in Figs.~\ref{ff1_fig} and~\ref{ff2_fig}.
}
\vspace{.4cm}
\begin{ruledtabular}
\begin{tabular}{c c c c c c c c c}
\vspace{-.2cm}
\hspace{.05cm} Fit \hspace{.05cm} & 
\hspace{.2cm} $L_4^r\:[10^{-3}]$ \hspace{.2cm} &
\hspace{.2cm} $L_5^r\:[10^{-3}]$ \hspace{.2cm} &
\hspace{.2cm} $L_6^r\:[10^{-3}]$ \hspace{.2cm} &
\hspace{.2cm} $L_8^r\:[10^{-3}]$ \hspace{.2cm} &
\hspace{.2cm} $\tilde C_\phi$ [keV$^{-\frac{1}{2}}$] \hspace{.2cm} & 
\hspace{.2cm} $\lambda_\phi$ \hspace{.2cm} &
\hspace{.2cm} $q_\mathrm{max}^{}$ [GeV] \hspace{.2cm} &
\hspace{.1cm} $\langle r^2\rangle_s^\pi$ [fm$^2$] \hspace{.1cm} \\
&&&&&&&& \\ \hline\hline
\vspace{-.1cm}
&&&&&&&& \\
\vspace{-.1cm}
I & 
$0.84\,^{+0.06}_{-0.05}$ & 
$0.45\,^{+0.08}_{-0.09}$ & 
$0.03\,^{+0.16}_{-0.13}$ & 
$0.33\,^{+0.14}_{-0.17}$ & 
$42.1\,^{+5.0}_{-5.0}$ & 
$0.132\,^{+0.018}_{-0.015}$ & 
($0.90 \pm 0.025$) & 
0.657 \\
&&&&&&&& \\
\vspace{-.1cm}
Ia &
$0.817\,^{+0.048}_{-0.029}$ & 
$0.384\,^{+0.009}_{-0.021}$ & 
$-0.025\,^{+0.110}_{-0.079}$ & 
$0.328\,^{+0.133}_{-0.172}$ & 
$45.27\,^{+1.82}_{-3.07}$ & 
$0.133\,^{+0.017}_{-0.016}$ & 
(0.875) & 
0.652 \\
&&&&&&&& \\
\vspace{-.1cm}
Ib &
$0.840\,^{+0.043}_{-0.023}$ & 
$0.450\,^{+0.006}_{-0.018}$ & 
$0.028\,^{+0.107}_{-0.075}$ & 
$0.325\,^{+0.129}_{-0.168}$ & 
$42.08\,^{+1.24}_{-2.49}$ & 
$0.132\,^{+0.016}_{-0.015}$ & 
(0.9) & 
0.657 \\
&&&&&&&& \\
\vspace{-.1cm}
Ic & 
$0.858\,^{+0.037}_{-0.022}$ & 
$0.520\,^{+0.006}_{-0.015}$ & 
$0.081\,^{+0.104}_{-0.078}$ & 
$0.323\,^{+0.134}_{-0.166}$ & 
$39.15\,^{+1.09}_{-2.03}$ & 
$0.131\,^{+0.016}_{-0.015}$ & 
(0.925) &
0.663 \\
&&&&&&&& \\ \hline
\vspace{-.1cm}
&&&&&&&& \\
\vspace{-.1cm}
II &
$0.95\,^{+0.08}_{-0.03}$ & 
$0.25\,^{+0.24}_{-0.29}$ & 
$0.06\,^{+0.13}_{-0.08}$ & 
$0.18\,^{+0.21}_{-0.25}$ & 
$44.7\,^{+1.2}_{-4.4}$ & 
$0.104\,^{+0.025}_{-0.023}$ &
($0.90 \pm 0.025$) & 
0.659 \\
&&&&&&&& \\ 
\vspace{-.1cm}
III &
$0.94\,^{+0.05}_{-0.04}$ & 
$0.40\,^{+0.10}_{-0.12}$ & 
($0.20$) & 
$0.12\,^{+0.11}_{-0.11}$ & 
$35.1\,^{+2.5}_{-2.6}$ & 
$0.123\,^{+0.015}_{-0.013}$ &
($0.90 \pm 0.025$) & 
0.670 \\
&&&&&&&& \\ 
\vspace{-.1cm}
IV &
$(0.70)$ & 
$0.55\,^{+0.12}_{-0.10}$ & 
$-0.15\,^{+0.11}_{-0.07}$ & 
$0.55\,^{+0.13}_{-0.14}$ & 
$50.9\,^{+3.8}_{-6.4}$ & 
$0.162\,^{+0.024}_{-0.017}$ &
($0.90 \pm 0.025$) & 
0.634 \\
&&&&&&&& \\ 
\end{tabular}
\end{ruledtabular}
\label{fit_tab}
\end{center}
\end{table*}

\begin{figure*}[t] 
\includegraphics[width=.85\textwidth]{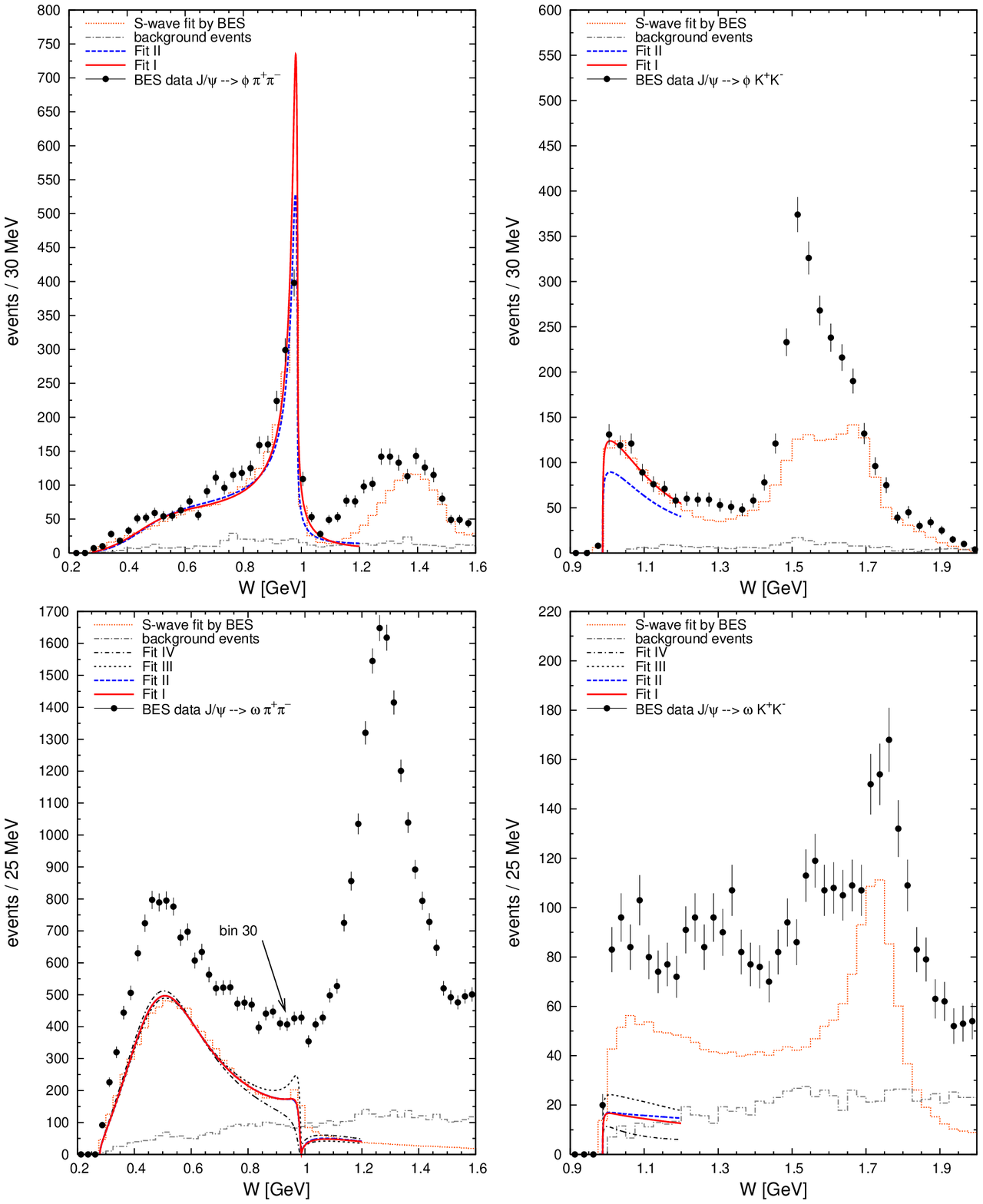} 
\caption{Fit results and BES data for $J/\psi \rightarrow 
\phi\pi^+\pi^-$, $J/\psi \rightarrow \phi K^+K^-$, $J/\psi \rightarrow 
\omega\pi^+\pi^-$ and $J/\psi \rightarrow \omega K^+K^-$. The 
datapoints are uncorrected for background and detection efficiency, and 
the errorbars represent the statistical error. The 
direct $S$-wave contribution from the BES PWA and the estimated 
background are denoted by the dashed and dotted histographs, 
respectively. The solid and dashed curves describe ``Fit I" and ``Fit 
II" to the $S$-wave contributions. As 
described in the text, ``Fit III" is performed for a fixed value of 
$L_6^r$ and ``Fit IV" for a fixed value of $L_4^r$. The fitted 
parameter values are given for each case in Table~\ref{fit_tab}. As 
discussed in Sect.~\ref{omg_res}, the $J/\psi \rightarrow \omega 
K^+K^-$ spectrum is not included in the fits.} 
\label{fit_fig} 
\end{figure*}

In order for the fit results to be well reproducible, the various 
constant parameters which enter the expressions for the decay 
amplitudes should be accurately defined. These parameters include the 
masses of the light pseudoscalar mesons, for which the current 
experimental values~\cite{PDG} have been used. These are $m_\pi = 
0.13957$~GeV, $m_K = 0.49368$~GeV and $m_\eta = 0.5478$~GeV, while the 
value $f_\pi = 0.0924$~GeV has been adopted for the pion decay 
constant. The physical masses of the charged pions and kaons 
have been used in order for the $\pi^+\pi^-$ and $K^+K^-$ thresholds to 
coincide with the physical ones. Also, the physical $\eta$ meson mass 
has been used rather than the one given by the GO~relation. The $\eta$ 
meson mass appears in relatively few places in the 
expressions, and checks on the fits have indicated that replacement of 
the physical $\eta$ mass with that given by the GO~relation has a 
minimal effect. Further parameters are the masses of the vector 
mesons $\rho,\omega,\phi$ and $J/\psi$. The $\rho$ mass is used in 
the evolution of the $L_i^r$, and has been taken as $m_\rho = 
0.776$~GeV. The other vector meson masses appear in various 
phase-space factors, and have been given the values $m_\omega = 
0.783$~GeV, $m_\phi = 1.020$~GeV and $m_{J/\psi} = 
3.097$~GeV~\cite{PDG}.

It is not {\it a priori} obvious how the individual deviations, 
required for the $\chi^2$ fit, are to be treated for the BES data. The 
individual deviations for each bin $\sigma_i$, are given by the BES 
collaboration as the square root of the number of events $\sqrt{N_i}$. 
These numbers represent the statistical errors of the raw data, 
uncorrected for detection efficiency and background. Furthermore, 
what is fitted in the present analysis is not the total signal detected 
by the BES experiment, but rather the $S$-wave contribution from the 
accompanying PWA. In view of these considerations, the 
deviations for each bin used in the fitting procedure have been taken 
to be of the form
\begin{equation}
\Delta_i \equiv \sqrt{N_i}\,w_i,
\end{equation}
where the $w_i$ represent weighting factors which have been chosen in a 
physically motivated way. In principle, individual $w_i$ could be 
introduced in all the decay channels studied and would then represent a 
``quality factor" for each bin. In practice, to avoid excessive 
fine-tuning of the fit, a constant value for $w_i$ has been applied for 
each decay channel, according to the following principles: 
Since the $S$-wave contribution in the $\omega\pi^+\pi^-$ spectrum is 
likely to have the largest uncertainty, a value of $w_i = 3$ has been 
adopted for that decay channel, whereas the values of $w_i$ for the 
$\phi\pi^+\pi^-$ and $\phi K^+K^-$ distributions have been defined to 
be between 0~and~1. The choice of such a large value of $w_i$ for 
$\omega\pi^+\pi^-$ means that the shape of the $\pi^+\pi^-$ spectrum is 
constrained by the $\phi\pi^+\pi^-$ and $\phi K^+K^-$ distributions, 
and may therefore be considered as a prediction of the present 
analysis. On the other hand, the relative magnitude of the 
$\phi\pi^+\pi^-$ and $\omega\pi^+\pi^-$ spectra provides an important 
piece of experimental input, without which all parameters in the fit 
cannot be constrained. Checks have been performed to verify that the 
choice of $w_i = 3$ for $\omega\pi^+\pi^-$ does not introduce 
unphysical skewing of the fit, and it was found that a further increase 
in $w_i$ for $\omega\pi^+\pi^-$ has a negligible effect on the fit 
results. The values for $w_i$ used for the other two channels are 
$w_i = 1$ for $\phi \pi^+\pi^-$ and $w_i = 0.1$ for $\phi K^+K^-$ in 
``Fit~I" and vice versa for ``Fit~II". Again, the fit results have been 
checked for stability with respect to these choices.

The predicted number of events $N_i$ in each 
bin is fitted to the total number of $S$-wave events determined by the 
BES PWA. To this end, the definition
\begin{equation}
\overline{N}(W) \equiv \left(\frac{dN}{dW}\right) \Delta W,
\label{nbar}
\end{equation}
has been implemented, where the differential event distribution $dN/dW$ 
is given by Eq.~(\ref{diffev}), and $\Delta W$ denotes the binsize 
chosen by BES for the PWA, given as 25~MeV for $\omega$~final states 
and 30~MeV for $\phi$~final states~\cite{BES1,BES2,BES3}. It should be 
reemphasized that all effects due to detection efficiencies are 
included in $dN/dW$ as defined in Eq.~(\ref{diffev}). The number of 
events $N_i$ per bin are then calculated by averaging Eq.~(\ref{nbar}) 
over the energy range $[W_1^i,W_2^i]$ spanned by the bin, giving
\begin{equation}
N_i = \frac{1}{\Delta W} \int_{W_1^i}^{W_2^i} \overline{N}(W)\:dW,
\end{equation}
where the explicit dependence on the binsize $\Delta W$ cancels. The 
advantage of the function $\overline{N}(W)$ is that it can be 
conveniently compared with the experimental data, regardless of the 
binsize chosen for a particular decay channel.

Finally, the $L_i^r$ determined in the fit are always quoted at the 
scale $\mu = m_\rho$. The fit function itself depends on the 
$L_i^r$ at the scale given by Eq.~(\ref{mu_match}) in terms of the 
cutoff parameter $q_\mathrm{max}$ of the FSI, which gives the $L_i^r$ 
at a scale of $\mu \sim 1.2 q_\mathrm{max}$. For calculational 
purposes, the $L_i^r$ are evolved to this scale.

\subsection{Error Analysis}
\label{fit_err}

All fits in this work are performed by minimizing the $\chi^2$ with 
respect to the $S$-wave event distributions reported by the BES 
PWA~\cite{BES1,BES2,BES3} for $\phi\pi^+\pi^-$, $\omega\pi^+\pi^-$ and 
$\phi K^+K^-$. The only parameter left fixed in all of the fits is the 
momentum cutoff $q_\mathrm{max}$, which is constrained by the 
requirement that the $f_0(980)$ resonance generated by the FSI should 
be well centered on the corresponding peak in the experimental 
$\phi\pi^+\pi^-$ spectrum. A rather broad $\chi^2$ minimum is found 
around a cutoff of $\sim 0.9$~GeV, and in order to avoid unnecessary 
fine-tuning, that value is taken as the central value to be used in all 
fits. To estimate the uncertainty associated with the choice of 
$q_\mathrm{max}$, a range of 50~MeV, between $q_\mathrm{max} = 
0.875$~GeV and $q_\mathrm{max} = 0.925$~GeV, has been considered. 
Larger or smaller values produce an $f_0(980)$ peak which is no longer 
well centered on the experimental one in $\phi\pi^+\pi^-$. In order to 
illustrate the dependence on the choice of cutoff $q_\mathrm{max}$, 
fits for a fixed cutoff, each corresponding to the values quoted above, 
have been given in Table~\ref{fit_tab} as \mbox{``Fit Ia - Ic"}. 

The statistical errors of the BES data~\cite{BES1,BES2,BES3} are small 
and have little bearing on the actual sources of uncertainties of the 
fitted parameters. The main sources of error accounted for in the 
present analysis are as follows: Firstly, the $\sigma$ amplitude used 
in the PWA for $\omega\pi^+\pi^-$ published by BES~\cite{BES2} did not 
take into account channel coupling effects, which leads to a 
considerable uncertainty in the $S$-wave contribution close to the 
$K\bar K$ threshold. In view of this, the $\omega\pi^+\pi^-$ data is 
only fitted up to bin~30 and the uncertainty on the parameter values 
introduced by this arbitrary choice is estimated via fits up to bins~29 
and~31. It is found that the effect of this is rather small for most 
parameters, and all of the fits contain a sharp characteristic cutoff 
in the $S$-wave $\pi^+\pi^-$ distribution close to the $K\bar K$ 
threshold. Secondly, the precise magnitude and shape of the $S$-wave 
contribution 
to the conspicuous low-energy enhancement, or $\sigma$ peak in the 
$\omega\pi^+\pi^-$ data has to be considered as somewhat uncertain, as 
the BES PWA models the $\sigma$ contribution in terms of 
different functional forms related to a relativistic Breit-Wigner 
description. Such approaches are known to suffer from consistency 
problems with respect to the data on $\pi\pi$ phase shifts, and indeed 
only one of the BES fits takes those data simultaneously into account. 
The $S$-wave contribution, as determined by the BES PWA constitutes 
about 60\% of the total signal in the $\sigma$ peak of the 
$\omega\pi^+\pi^-$ spectrum. The fact that the properties of 
the $\sigma$ peak are strongly influenced by interference with 
rescattering effects in the $\omega\pi$ system has also been 
demonstrated in Ref.~\cite{Roca}, where a sequential decay mechanism 
with intermediate vector and axial-vector mesons, e.g. $J/\psi 
\rightarrow b_1(1235)\pi \rightarrow \omega\pi\pi$ was investigated.

The situation in the BES PWA is similar, as most of the 
remaining events at the $\sigma$ peak are due to the interference of 
the $\sigma$ amplitude with the prominent $b_1(1235)$ and $\rho(1450)$ 
resonances in the $\omega\pi$ system. At higher energies, the left wing 
of the strong $D$-wave $f_2$(1270) resonance also contributes 
significantly. The situation for the $\phi\pi^+\pi^-$ spectrum is much 
more straightforward since the $S$-wave contribution completely 
dominates the data up to the $K\bar K$ threshold, and also because of 
the much weaker interaction in the $\phi\pi$ system. The advantage of 
using the BES PWA is that the direct $S$-wave contribution to the 
$\sigma$ peak has been disentangled from the $b_1(1235)$ and 
$\rho(1450)$ contributions, such that a direct fit to the $S$-wave 
spectrum is possible without consideration of strong rescattering 
effects in the $\omega\pi$ system, which would introduce further 
uncertainties due to the modeling of such effects. The downside 
is that such uncertainties are included instead in the determination of 
the $S$-wave amplitude in the PWA.

In view of the abovementioned issues, a conservative $10\%$ 
uncertainty in the size of the $S$-wave contribution has been adopted 
for the $\omega\pi^+\pi^-$ spectrum to estimate the errors on the 
fitted parameter values. Moreover, as described in Sect.~\ref{fit_def}, 
the $\omega\pi^+\pi^-$ data has been given a rather low weight factor, 
and thus the shapes of the fitted spectra are essentially dictated by 
the $\phi\pi^+\pi^-$ and $\phi K^+K^-$ channels. Nevertheless, 
inclusion of the $\omega\pi^+\pi^-$ data in the fits is essential in 
order to provide information on the relative intensities of the 
$\phi\pi^+\pi^-$ and $\omega\pi^+\pi^-$ channels. Without that 
information, the model parameter $\tilde C_\phi$ cannot be properly 
constrained since the present BES data~\cite{BES1,BES2,BES3} are 
unnormalized, which was also the case in Ref.~\cite{UGM1}.

\subsection{Fit Results for $\phi\pi^+\pi^-$ and $\phi K^+K^-$}
\label{phi_res}

The experimental dipion spectrum for $J/\psi \rightarrow 
\phi\pi^+\pi^-$ is dominated by the conspicuous $S$-wave $f_0(980)$ 
resonance, such that an accurate description of that state is necessary 
in order for a reasonable fit to be attained. In the present 
description, the $f_0(980)$ is dynamically generated and its appearance 
is a consequence of the channel coupling to $K\bar K$. However, 
since the FSI model depends on only one parameter, namely the momentum 
cutoff $q_\mathrm{max}$, the possibilities of optimizing the width of 
that state are severely restricted. It is thus fortunate that the 
present considerations give a reasonable description of the $f_0(980)$, 
with the possible objection of that state being slightly too narrow. 
This generates an ambiguity which is reflected in the fact that two 
different fits, shown as ``Fit~I" and ``Fit~II" in Fig.~\ref{fit_fig}, 
may be achieved depending on whether the $\phi\pi^+\pi^-$ or the $\phi 
K^+K^-$ spectrum is optimized. 

As in the present case for ``Fit~II", agreement was optimized for 
$\phi\pi^+\pi^-$ in Ref.~\cite{UGM1}, which resulted in a noticeable 
underprediction of the $\phi K^+K^-$ spectrum. As explained in 
Sect.~\ref{fit_def}, ``Fit~I" was defined by forcing agreement with the 
$\phi K^+K^-$ spectrum, which may be accomplished at the small price of 
a moderate overprediction in the $\phi\pi^+\pi^-$ spectrum, which 
amounts to a factor~$\sim 2$ at the bin centered on the $f_0(980)$ 
peak. A comparison of the fitted parameter values given in 
Table~\ref{fit_tab} suggests that the ones corresponding to ``Fit~I" 
are more realistic. Further comparison of the two fits has also 
revealed that ``Fit~I" is more stable against variations in the input. 
It is thus concluded that ``Fit~I" is preferable and that the modest 
overprediction at the $f_0(980)$ resonance in the $\phi\pi^+\pi^-$ 
spectrum is due to the limitations of the FSI model employed in this 
paper. It should be emphasized that this discrepancy is small enough to 
go unnoticed in a study of meson-meson phase shifts alone. Also, the 
difference between ``Fit~I" and ``Fit~II" is negligible in the 
$\omega\pi^+\pi^-$ spectrum.

Finally, the observed excess of events in the left wing of the 
$f_0(980)$ resonance in the $\phi\pi^+\pi^-$ spectrum should be 
mentioned, which indicates a significant presence of a nonstrange, or 
$\sigma$ contribution. Indeed, the fitted value of $\lambda_\phi \sim 
0.13$ given in Table~\ref{fit_tab} reproduces this feature well.

\subsection{Fit Results for $\omega\pi^+\pi^-$ and $\omega K^+K^-$}
\label{omg_res}

\begin{figure*}[t]
\includegraphics[width=.85\textwidth]{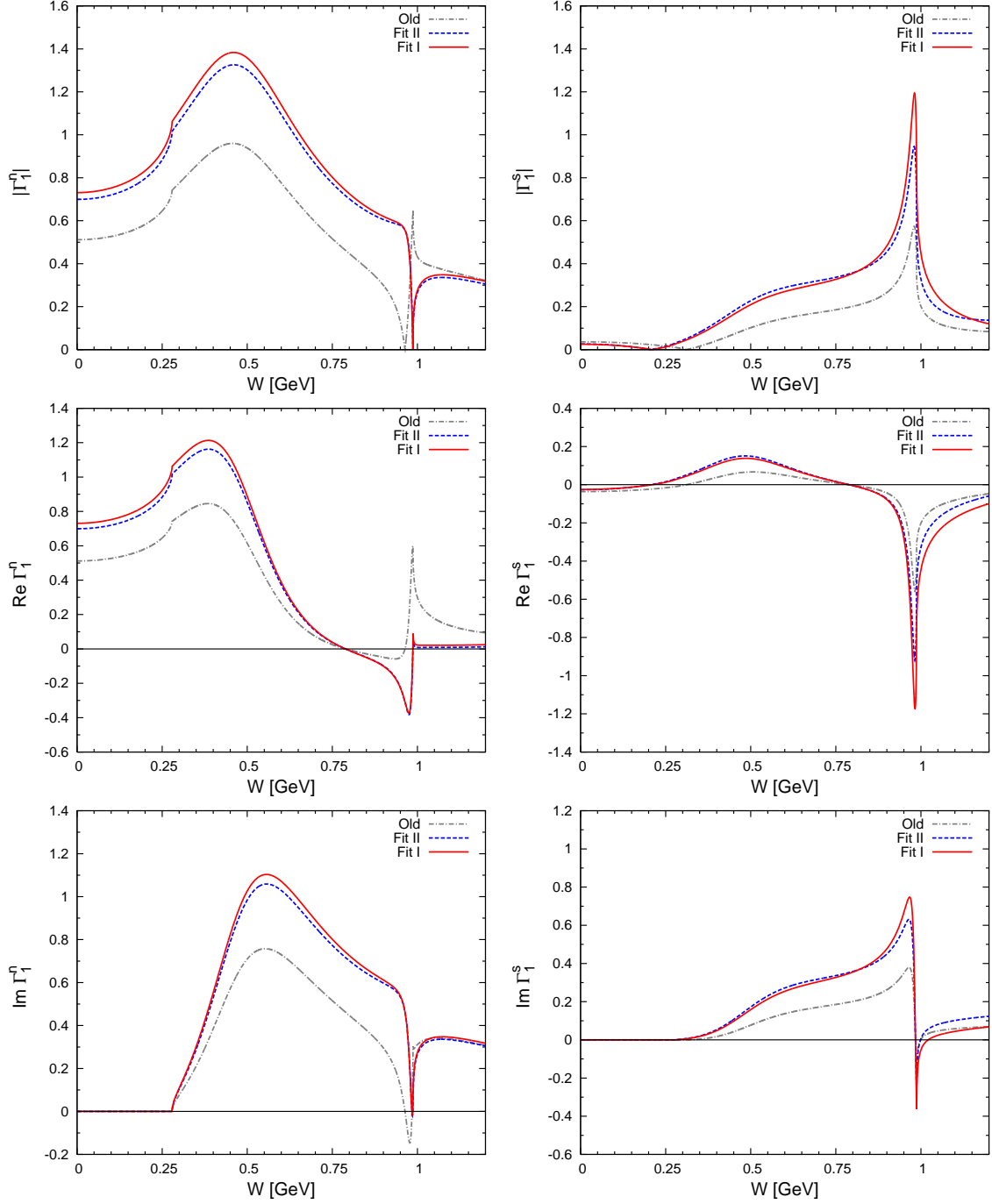}
\caption{The nonstrange and strange scalar~FFs $\Gamma_1^n$ 
and $\Gamma_1^s$ of the pion. Moduli, real and imaginary parts are 
shown for the parameters corresponding to ``Fit I" and ``Fit II" in 
Table~\ref{fit_tab}. The curves labeled ``Old" show, for comparison, 
the scalar~FFs for the parameters obtained in the analysis of 
Ref.~\cite{UGM1}.}
\label{ff1_fig}
\end{figure*}

The most conspicuous features in $\omega\pi^+\pi^-$ are the lack of 
a significant peak at the position of the $f_0(980)$ 
resonance, and a very prominent $\sigma$ peak around a dipion energy 
of $\sim 0.5$~GeV. It is evident that ``Fit~I" in the lower left-hand 
plot in Fig.~\ref{fit_fig} faithfully reproduces both of these 
features. The $\sigma$ peak is generated by the dominance of the pion 
nonstrange~FF, which in contrast to the strange one approaches a 
constant value at low energies. The main difference between the 
$S$-wave contribution of the BES PWA and the present fit is 
close to the $K\bar K$ threshold, where the fitted spectrum is sharply 
cut off by channel-coupling effects, producing a plateaulike 
appearance close to the $f_0(980)$ resonance. This characteristic shape 
is produced here by the interplay of the strange and nonstrange 
scalar~FFs close to the $K\bar K$ threshold. It should also be noted 
that most of the remaining $\pi\pi$ events below the $K\bar K$ 
threshold are, as elaborated in Sect.~\ref{fit_err}, due to 
interference effects with the $b_1(1235)$ and $\rho(1450)$ resonances 
in the $\omega\pi$ system.

Once the parameters $L_i^r$, $\tilde C_\phi$ and $\lambda_\phi$ are 
fixed by a fit to the $\phi\pi^+\pi^-$, $\omega\pi^+\pi^-$ and $\phi 
K^+K^-$ spectra, the scalar~FFs can be used to predict the $S$-wave 
contribution in the $\omega K^+K^-$ channel. It turns out that such a 
prediction is quite robust, since attempts at tweaking the fit 
by forcing changes in the $\omega K^+K^-$ sector do not have any 
appreciable effect. Nevertheless, it is immediately apparent from 
Fig.~\ref{fit_fig} that the $S$-wave contribution reported by the BES 
collaboration does not compare favorably with the present results. The 
$S$-wave fitted by BES contributes $\sim 50$~events close to the $K\bar 
K$ threshold, whereas the present prediction can account for $\sim 
20$~events. This number can be motivated from the present 
considerations as follows: From Figs.~\ref{ff1_fig} and~\ref{ff2_fig}, 
the moduli of the nonstrange pion and kaon~FFs are seen to be 
approximately equal in magnitude, such that a roughly similar $S$-wave 
contribution would be expected at $\sim 1.1$~GeV in the 
$\omega\pi^+\pi^-$ and $\omega K^+K^-$ spectra. This would lead one to 
expect $\sim 50$ events in the $\omega K^+K^-$ spectrum as well, 
however the detection efficiency reported by BES for $\omega K^+K^-$ is 
only about half of that reported for $\omega\pi^+\pi^-$, yielding the 
calculated $\sim 20$~events. The situation in the $\phi K^+K^-$ 
spectrum is different, as the modulus of the strange kaon~FF 
at $\sim 1.1$~GeV is much larger than that of the strange pion~FF, as 
evidenced by Figs.~\ref{ff1_fig} and~\ref{ff2_fig}. 

With respect to this discrepancy, it should be noted that when the BES 
PWA was done, no reliable information was available on the 
magnitude of a possible $\sigma$ coupling to $K\bar K$, and therefore 
it was fitted freely in that analysis. The BES data exhibits 
interference between $\sigma \rightarrow K\bar K$ and a signal fitted 
as $K_1(1945) \rightarrow \omega K$. It turns out that the latter can 
be adjusted slightly in mass and width in order to reduce the $\sigma 
\rightarrow K\bar K$ signal to a level consistent, within statistical 
and systematical errors, with the findings of the analysis presented in 
this paper, at the price of a slightly worse fit to the $K\bar K$ event 
distribution\footnote{The authors thank David Bugg
for his kind cooperation in investigating the discrepancy between the 
predicted $\omega K^+K^-$ spectrum and the BES analysis.}. The 
overall changes induced in the BES PWA by such modifications are very 
small.

\subsection{Analysis of the $L_i^r$}

One of the main objectives of this analysis is the determination of the 
LECs $L_4^r$ and $L_6^r$, for which the data + PWA provided by 
the BES collaboration is certainly precise enough. As discussed in 
Refs.~\cite{GL1,DAPHNE}, where a set of standard estimates for the 
LECs of CHPT are given, $L_4^r$ and $L_6^r$ are expected, in terms of 
large $N_c$ arguments, to vanish at some unknown scale in the 
resonance region of QCD, conventionally taken to be $m_\rho$ or 
$m_\eta$. As discussed above, ``Fit~I", as given in 
Table~\ref{fit_tab} is considered the main result in this respect. The 
values so obtained, $(0.84 \pm 0.06) \times 10^{-3}$ for $L_4^r$ and 
$(0.03 \pm 0.16) \times 10^{-3}$ for $L_6^r$ should be compared with 
the NNLO CHPT analyses of Refs.~\cite{HB1a,HB1b,HB2}, where the $L_i^r$ 
were extracted by fits to the available experimental data on $K_{l4}$ 
decays. In that study, constraints on $L_4^r$ and $L_6^r$ were also 
derived by requiring a properly convergent behavior of the chiral 
expansion up to NNLO for several quantities, such as the pseudoscalar 
meson masses and decay constants. It was found that such a constraint 
can only be satisfied in a rather small region centered around 
$L_4^r = 0.2 \times 10^{-3}$ and $L_6^r = 0.5 \times 10^{-3}$. Also of 
interest in this context is the analysis based on QCD sum rules in 
Ref.~\cite{MSS1}, where $L_6^r$ was constrained to be positive and in 
the region $0.2 \times 10^{-3} \leq L_6^r \leq 0.6 \times 10^{-3}$. 
Also, a significantly positive value of $L_4^r = 0.4 \times 10^{-3}$ 
was obtained in Ref.~\cite{MSS1}, although with unknown error. Some 
updated results for the $L_i^r$ are given in Ref.~\cite{HB4}, and a 
recent analysis of $\pi\pi$ and $\pi K$ scattering is 
presented in Ref.~\cite{HB5}, where the preferred values of $L_4^r$ and 
$L_6^r$ were found to be compatible with zero. Further recent 
determinations of the $L_i^r$ include the analyses of 
Refs.~\cite{Pel1,Pel2} and~\cite{Guerr} in terms of the Inverse 
Amplitude Method (IAM) of Ref.~\cite{IAM}. In Ref.~\cite{Guerr}, a 
simultaneous fit to the $\pi\pi$ and $K\bar K$ partial wave amplitudes 
for $I=0,1,2$ was performed using the complete NLO CHPT amplitudes, 
which yielded a value 
of $L_4^r$ equal to $(0.2 \pm 0.1) \times 10^{-3}$. The analysis of 
Ref.~\cite{Pel2} considered all two-meson scattering amplitudes 
including the $\eta\eta$ channel within the chiral IAM framework, which 
enabled the simultaneous extraction of the LECs $L_1^r$ through 
$L_8^r$, the reported values of $L_4^r$ and $L_6^r$ being $(-0.36 \pm 
0.17) \times 10^{-3}$ and $(0.07 \pm 0.08) \times 10^{-3}$, 
respectively. Finally, Lattice QCD also offers the possibility of 
determining the LECs of CHPT via application of Partially Quenched~(PQ) 
Lattice QCD and PQCHPT. For a theoretical background, see 
Refs.~\cite{QQCD,PQQCD,BDL,BL,Latt}, and some recent Lattice QCD 
results for the $L_i^r$ can be found in Refs.~\cite{Alpha,qqq,MILC}.

\subsection{Results for $L_4^r$}

\begin{figure*}[t]
\includegraphics[width=.85\textwidth]{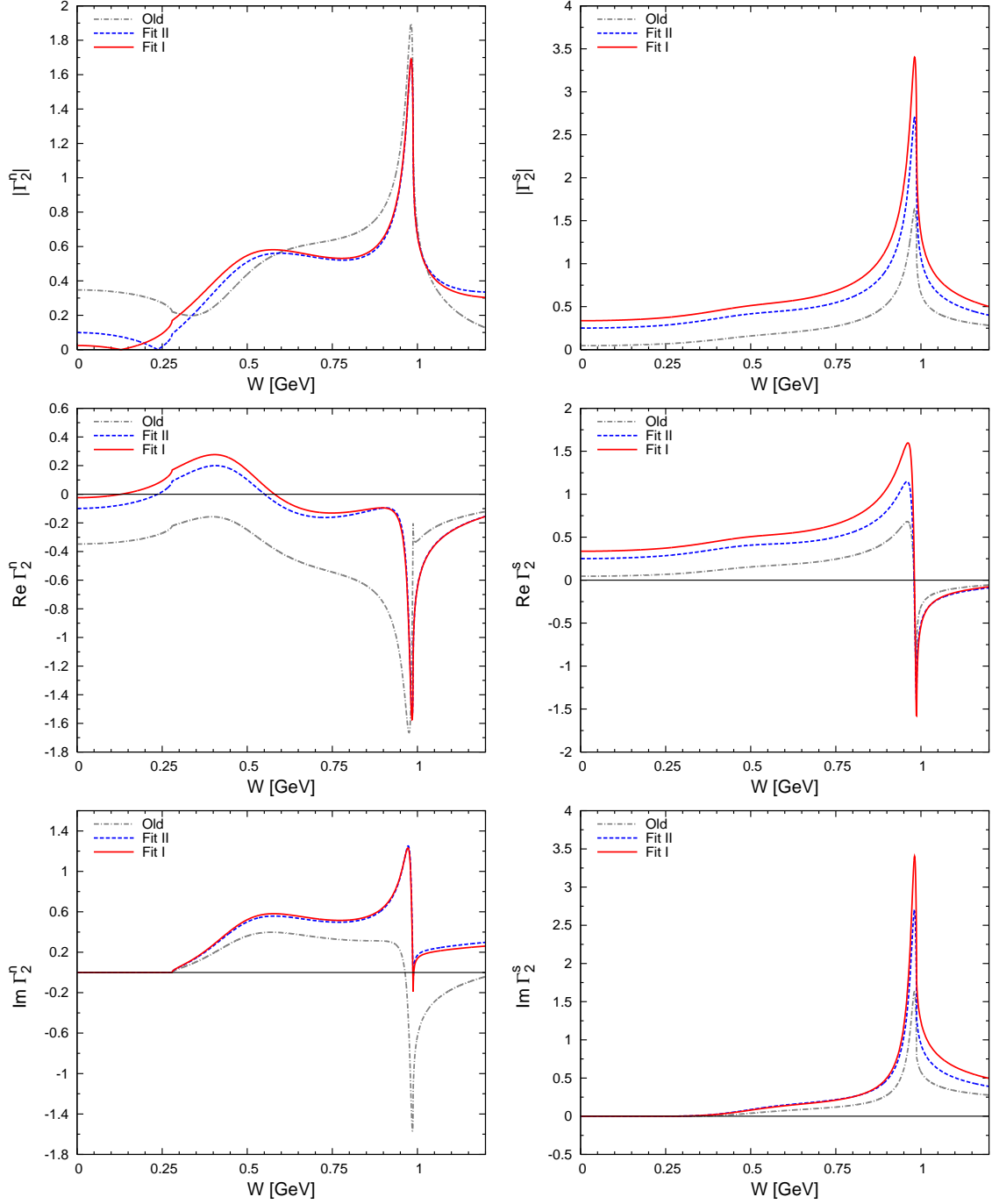}
\caption{The nonstrange and strange scalar~FFs $\Gamma_2^n$ 
and $\Gamma_2^s$ of the kaon. Moduli, real and imaginary parts are 
shown for the parameters corresponding to ``Fit I" and ``Fit II" in 
Table~\ref{fit_tab}. The curves labeled ``Old" show, for comparison, 
the~FFs for the parameters obtained in the analysis of 
Ref.~\cite{UGM1}.}
\label{ff2_fig}
\end{figure*}

The value of $L_4^r$ preferred by the present study is $(0.84 \pm 
0.06) \times 10^{-3}$, which is largely consistent with the value 
$(0.71 \pm 0.11) \times 10^{-3}$ reported in Ref.~\cite{UGM1} where a 
similar analysis was performed, although with much poorer data and 
without the guidance of a comprehensive PWA. These results are about 
twice as large as the largest values found by the other analyses 
presented above. Also, the present result for $L_4^r$ is difficult to 
reconcile with the IAM analysis of 
Ref.~\cite{Pel2}, where a rather large but negative value of $L_4^r 
\sim -0.4 \times 10^{-3}$ was obtained. However, the results of that 
IAM analysis are not very sensitive to the value of $L_4^r$, since it 
has been pointed out in Ref.~\cite{Pel3} that values of $L_4^r \sim 0$ 
or $L_4^r \sim 0.2 \times 10^{-3}$ can be accommodated with small 
overall changes.

The present value of $L_4^r$ is constrained by a simultaneous fit to 
the $\phi\pi^+\pi^-$ and $\omega\pi^+\pi^-$ spectra. Even a modest 
deviation from the central value of $L_4^r$ tends to upset the 
interference between the strange and nonstrange~FFs at the $f_0(980)$ 
peak, producing for large deviations a pronounced cusp in the 
$\omega\pi^+\pi^-$ spectrum which contradicts the flat, plateaulike 
behavior seen in the experimental data. Nevertheless, the 
$\omega\pi^+\pi^-$ spectrum is not pathologically sensitive to the 
exact properties of the $f_0(980)$ resonance, 
since the bins centered on that energy in $\omega\pi^+\pi^-$ have been 
excluded from the fit; Rather, a smaller value of $L_4^r$ as used for 
``Fit~IV", starts to upset the behavior of the low-energy parts of both 
the $\phi\pi^+\pi^-$ and $\omega\pi^+\pi^-$ spectra. It is thus 
encouraging to note that once $L_4^r$ is allowed to adjust itself 
optimally such that the low-energy region is well described, then the 
$\omega\pi^+\pi^-$ spectrum also ``finds" the right shape at the 
$K^+K^-$ threshold. The conclusion is therefore that a smaller value of 
$L_4^r$ cannot easily be accommodated within the present analysis. 
Since the OZI rule is also based on large $N_c$ arguments, a large 
value of $L_4^r$ may be interpreted as a clear signal of OZI violation. 
However, it should be emphasized here that the inclusion of 
higher-order terms in the vertex of Eq.~(\ref{Lagr}) may affect the value of 
$L_4^r$ significantly, although as discussed in Sect.~\ref{ff_res} 
there is little evidence as yet for such higher-order contributions.

\subsection{Results for $L_6^r$}

For $L_6^r$, the value $(0.03 \pm 0.16) \times 10^{-3}$ determined from 
``Fit~I" is in much better agreement with the results of the other 
analyses above. In remarkable contrast to the situation for $L_4^r$, 
the present value of $L_6^r$ is completely consistent with that 
determined by the IAM method in Ref.~\cite{Pel2}. Although the present 
result is consistent with zero within errors, a fixed 
value of around $0.2 \times 10^{-3}$, as suggested by the bounds on 
$L_6^r$ derived in Ref.~\cite{MSS1}, does little to upset the agreement 
with experiment, as indicated by ``Fit~III", the main effect being a 
slight overprediction close to the $f_0(980)$ resonance in the 
$\omega\pi^+\pi^-$ spectrum. 

As already noted in Sect.~\ref{phi_res}, 
the $f_0(980)$ as generated by the FSI is slightly too pronounced in 
the present framework, which may be interpreted as pulling the fit 
toward lower values of $L_6^r$. Anyway, as the main sensitivity to 
$L_6^r$ lies in the region of the $f_0(980)$ resonance, it is 
unfortunate that the presently available version of the BES PWA 
does not take into account channel coupling effects for that state. If 
the bins of $\omega\pi^+\pi^-$ data around the $f_0(980)$ could be 
included in the fit, the uncertainty on $L_6^r$ would most likely be 
much reduced. The value of $L_6^r$ found by the old analysis in 
Ref.~\cite{UGM1} is $-0.22 \times 10^{-3}$, without any error 
specified, which is inconsistent with the lower bound given in 
Ref.~\cite{MSS1}. However, the basic reason such a value was obtained 
is that the $\omega\pi^+\pi^-$ spectrum was assumed, due to the lack of 
a PWA, to be completely dominated by the $S$-wave contribution 
at low energies. However, this may only constitute about $\sim 
60\%$ of the total event count, as indicated by the recent BES 
PWA. For this reason, the $\omega\pi^+\pi^-$ spectrum as given 
in Ref.~\cite{UGM1} does not have the proper shape close to the $K\bar 
K$ threshold.

\subsection{Results for $L_5^r$ and $L_8^r$}

The values of $L_5^r$ and $L_8^r$ are much less controversial, as most 
studies are in reasonable agreement concerning their numerical values. 
In the previous work of Ref.~\cite{UGM1} their values were kept fixed 
and were taken to coincide with those determined by the $K_{l4}$ 
analysis of Ref.~\cite{HB2}. These values were quoted as $L_5^r = (0.65 
\pm 0.12) \times 10^{-3}$ and $L_8^r = (0.48 \pm 0.18) \times 10^{-3}$, 
respectively. However, in the present analysis it turns out to be 
possible to let these parameters adjust themselves to their optimal 
values. From ``Fit~I" presented in Table~\ref{fit_tab}, the obtained 
values are $L_5^r = (0.45 \pm 0.09) \times 10^{-3}$ and $L_8^r = (0.33 
\pm 0.17) \times 10^{-3}$. Although they come out slightly smaller than 
the values of Ref.~\cite{HB2}, it is encouraging to note that an 
optimal fit to data does not require extreme values of $L_5^r$ and 
$L_8^r$. The more recent fit in Ref.~\cite{HB4} has 
reported an updated set of LECs which compare slightly more unfavorably 
with the present values, namely $L_5^r = (0.91 \pm 0.15) \times 
10^{-3}$ and $L_8^r = (0.62 \pm 0.20) \times 10^{-3}$. However, it 
should be kept in mind that the nature of the fits in Ref.~\cite{HB4} 
did not allow for a detailed consideration of the effects of the LECs 
$L_4^r$ and $L_6^r$. It should also be noted that the values for 
$L_5^r$ and $L_8^r$ obtained in Refs.~\cite{HB2,HB4} are significantly 
smaller, and therefore in much better agreement with the present 
results, than the old standard values $L_5^r = (1.4 \pm 0.5) \times 
10^{-3}$ and $L_8^r = (0.9 \pm 0.3) \times 10^{-3}$ in 
Ref.~\cite{DAPHNE}. Finally, it should be noted that the results of the 
IAM analysis of Ref.~\cite{Pel2} for $L_5^r$ and $L_8^r$ are 
essentially identical to the standard ones given in Ref.~\cite{DAPHNE}.

\subsection{Fitted Values for $\tilde C_\phi$ and $\lambda_\phi$}
\label{lphi_res}

The overall normalization constant $\tilde C_\phi$ is determined by the 
present analysis to be around $40$~keV$^{-\frac{1}{2}}$, as indicated 
in Table~\ref{fit_tab}. This value is difficult to compare with 
Ref.~\cite{UGM1}, as that work did not consider the detection 
efficiencies of the DM2 and MARK-III experiments. Furthermore, 
Ref.~\cite{UGM1} also employed a different definition of the overall 
normalization factor. In the present study, the dimension of $\tilde 
C_\phi$ is a consequence of the absorption of the proportionality 
factor in Eq.~(\ref{diffev}) into the product $\sqrt{2}\,gB_0\,C_\phi$. 
This factor could be estimated once a conversion of the 
event distributions reported by the BES collaboration to differential 
branching fractions is available. However, such an estimate is 
complicated by the nontriviality of the detection efficiencies shown in 
Fig.~\ref{eff_fig}, and by a considerable uncertainty in the hitherto 
published values for the relevant branching fractions~\cite{PDG}. In 
principle, $\tilde C_\phi$ could then be used, as in Ref.~\cite{Roca} 
to extract a value for the dimensionless product $gB_0$. A 
determination of the coupling constant $g$ itself is more difficult 
since the precise value of $B_0$, which is related to the quark 
condensate, is only approximately known. In view of the above issues, 
further analysis of $\tilde C_\phi$ is beyond the scope of this paper.

Of considerably more interest is the fitted value of the parameter 
$\lambda_\phi$, which directly measures the OZI violation in the 
$J/\psi \rightarrow \phi\pi^+\pi^-$ and related decays. From ``Fit~I" 
in Table~\ref{fit_tab}, the present value is $\lambda_\phi = 0.13\pm 
0.02$, which is lower than the value $0.17\pm 0.06$ 
obtained in Ref.~\cite{UGM1}, although both values are consistent given 
the rather large errors of the latter result. The reason for this 
apparent decrease in $\lambda_\phi$ can be immediately understood in 
terms of Eq.~(\ref{phirel}), if one notes that all of the events in the 
low-energy part of the $\omega\pi^+\pi^-$ spectrum were taken in 
Ref.~\cite{UGM1} to be $S$-wave events. Put another way, the $S$-wave 
contribution for $\omega\pi^+\pi^-$ was taken to be about $50\%$ larger 
in Ref.~\cite{UGM1} than it was found to be in the BES 
PWA. A larger value of $\tilde C_\omega$ was thus required, in turn 
yielding a larger value of $\lambda_\phi$ via Eq.~(\ref{phirel}). 
Nevertheless, the present value of $\lambda_\phi$ indicates a 
significant amount of OZI violation, which manifests itself as a 
contribution to $\phi\pi^+\pi^-$ from the nonstrange scalar~FF, and a 
contribution to $\omega\pi^+\pi^-$ from the strange one. In particular, 
a realistic description of the $\omega\pi^+\pi^-$ spectrum close to the 
$K\bar K$ threshold requires an interplay between the nonstrange and 
strange scalar~FFs.

In this context, it is useful to recall Ref.~\cite{Roca}, where the 
experimental results of Refs.~\cite{DM2,MK3,BES0} were normalized using 
the available information~\cite{PDG} on the width and branching 
fractions of the $J/\psi$. Since Ref.~\cite{Roca} considered only the 
LO~CHPT contributions to the operators $\bar nn$ and $\bar ss$, the 
values of the $L_i^r$ do not appear as additional parameters in that 
model. It should be pointed out though, that the FSI between pairs of 
pions and kaons was taken into account in a similar way as in 
Ref.~\cite{UGM1} and in the present work. Furthermore, the strength of 
the sequential decay mechanism via intermediate vector and axial-vector 
resonances could be fixed from experimental information~\cite{PDG} up 
to a relative sign of the respective coupling constants. Up to this 
relative sign, the only free parameters in the calculation of 
Ref.~\cite{Roca} were $gB_0$ and $\lambda_\phi$. The sign ambiguity 
could not be resolved in Ref.~\cite{Roca}, since equally good fits to 
the data were obtained for both choices. If both the $J/\psi \rightarrow 
VP$ and $J/\psi \rightarrow AP$ coupling constants have the same sign, 
Ref.~\cite{Roca} finds $gB_0 = 0.032 \pm 0.001$ and $\lambda_\phi = 
0.12 \pm 0.03$. This value of $\lambda_\phi$ is consistent with the one 
obtained from ``Fit I" in Table~\ref{fit_tab}.

\subsection{Results for the Scalar Form Factors}
\label{ff_res}

Once all parameters are determined by the analysis of the BES 
data~\cite{BES1,BES2,BES3}, the shapes and magnitudes of the scalar~FFs 
of the pion and the kaon are fixed. In Figs.~\ref{ff1_fig} 
and~\ref{ff2_fig}, they are shown for the fitted parameters given in 
Table~\ref{fit_tab} and compared to the form factors obtained in 
Ref.~\cite{UGM1}. It is evident that the values of the form factors at 
the origin have changed significantly with respect to those of 
Ref.~\cite{UGM1}, even though the gross features remain similar. The 
most significant change is that the form factors now have a realistic 
behavior close to the $K\bar K$ threshold, which is reflected in the 
values of the form factors at the origin. Thus $\Gamma_1^n(0)$ has 
increased by $\sim 30$\%, while the value of $\Gamma_1^s(0)$ remains 
very small and negative. The situation for the kaon form factors is 
more drastic, since they were not very well determined in 
Ref.~\cite{UGM1}. The present results indicate that $\Gamma_2^n(0)$ is 
small and negative, while $\Gamma_2^s(0)$ is positive and has increased 
significantly from the value determined by previous work.

Since the description of the experimental data shown in 
Fig.~\ref{fit_fig} is satisfactory, a natural question to ask is 
whether the scalar~FFs can be considered realistic as well. The issue 
is complicated by the essentially unknown momentum dependence of the 
vertex given in Eq.~(\ref{Lagr}), for which the simplest possible form 
was assumed in Ref.~\cite{UGM1}. The question is then whether the 
scalar~FFs are simulating some neglected momentum-dependence in the 
$J/\psi \rightarrow \phi$ and $J/\psi \rightarrow \omega$ vertices. 
Although the detailed study of this is beyond the scope of the present 
study, the scalar radius $\langle r^2\rangle_s^\pi$ of the pion and the 
curvature $c_s^\pi$ of the nonstrange scalar FF may nevertheless be 
used to check for consistency with other theoretical descriptions. 
These quantities are defined~\cite{GL2,GM1} in terms of the expansion
\begin{eqnarray}
\frac{\Gamma_\pi^n(s)}{\Gamma_\pi^n(0)} &=& 1 +
\frac{1}{6} \langle r^2\rangle_s^\pi s + c_s^\pi s^2 \:+\: 
\mathcal{O}(s^3),
\label{srad}
\end{eqnarray}
of the nonstrange scalar FF of the pion around $s=0$. The 
values for $\langle r^2\rangle_s^\pi$ given by ``Fit I" in 
Table~\ref{fit_tab} are quite stable around $0.657 \pm 0.006$~fm$^2$, 
whereas the values for $c_s$ are $\sim 12.50$~GeV$^{-4}$ for ``Fit I" 
and $\sim 12.53$~GeV$^{-4}$ for ``Fit II". These values can be compared 
with other theoretical predictions, starting from the prediction of 
Ref.~\cite{GL1}, where an estimate of $\langle r^2\rangle_s^\pi$ = 
$0.55 \pm 0.15$~fm$^2$ was obtained from the experimental value of 
$F_K/F_\pi$, neglecting OZI-violating contributions. This estimate has 
been refined by the dispersive work of Ref.~\cite{Donogh} and further 
improved upon in Ref.~\cite{CGL} using the $\pi\pi$ Roy equations to 
yield $\langle r^2\rangle_s^\pi$ = $0.61 \pm 0.04$~fm$^2$. In 
Ref.~\cite{Donogh}, the curvature $c_s^\pi$ was also determined and 
found to be around $\sim 10.6$~GeV$^{-4}$, which is about $10\%$ 
smaller than the present value. In the NNLO CHPT analysis of 
Ref.~\cite{HB1b}, the value of $c_s^\pi$ could not be pinned down very 
precisely, although the preferred values are in the range 
$10 \ldots 13$~GeV$^{-4}$.

The effects of inelasticities due to the $\pi\pi \rightarrow 4\pi$ and 
$\pi\pi \rightarrow \eta\eta$ channels were studied in detail in 
Ref.~\cite{MSS1}, which produced a range of values for $\langle 
r^2\rangle_s^\pi$ consistent with the result of Ref.~\cite{CGL}. 
Higher-order CHPT corrections to the scalar FFs have been considered in 
Ref.~\cite{GM1} and in the NNLO fits of Refs.~\cite{HB1a,HB1b}, which 
also yielded a preferred value of $\langle r^2\rangle_s^\pi$ around 
$\sim 0.61$~fm$^2$. Finally, the value of $\langle r^2\rangle_s^\pi$ 
has also been extracted from the $S$-wave $\pi\pi$ phase shift for 
$I=0$ using the sum rule of Ref.~\cite{YND1}, which yielded a value of 
$0.75 \pm 0.07$~fm$^2$. However, as stated in Ref.~\cite{Ananth}, this 
work is not in agreement with earlier determinations of $\langle 
r^2\rangle_s^\pi$.

The conclusion of this comparison is that the present values for the 
pion scalar radius $\langle r^2\rangle_s^\pi$ and the curvature 
$c_s^\pi$ of the pion scalar~FF are largely consistent with other 
recent theoretical descriptions. The present $\langle r^2\rangle_s^\pi$ 
value of $\sim 0.65$~fm$^2$ is at the upper limit of the generally accepted 
value of $0.61 \pm 0.04$~fm$^2$. It is possible that the slightly 
higher value found in the present study is hinting at a neglected 
momentum dependence in the vertex of Eq.~(\ref{Lagr}) or at some 
inaccuracy in the present description of the $\pi\pi$ phase shifts. 
Nevertheless, the reasonable value of $\langle r^2\rangle_s^\pi$ is 
reassuring in view of the somewhat unorthodox value of $L_4^r$ obtained 
from the present analysis.

\section{Discussion and Outlook}
\label{concl}

The main results of this analysis of the experimental BES 
results~\cite{BES1,BES2,BES3} are the set of values obtained for the 
$L_i^r$ and the corresponding sets of scalar~FFs of the pions and 
kaons. Due to the accuracy of the BES data, the actual errors of the 
$L_i^r$ are dominated by the uncertainty in the theoretical 
description. The present value of $L_6^r$ is small and compares well 
with the known theoretical bounds. The present value is compatible with 
zero, although on the positive side. On the other hand, $L_4^r$ was 
found to be large and positive, around $(0.84 \pm 0.06) \times 
10^{-3}$. It has also been shown that a value of $L_4^r$ smaller than 
about $0.7\times 10^{-3}$ is not compatible with the present analysis. 
Although this value of $L_4^r$ is in conflict with most of the 
presently available fits of the $L_i^r$, an advantage of the present 
analysis is that the expressions for the scalar~FFs depend on only four 
LECs, namely $L_4^r$,$L_5^r$,$L_6^r$ and $L_8^r$, and that no 
assumptions about their values had to be made, since all of them 
could be constrained in a unique way. Also, it should be kept in mind 
that the present description of the scalar FFs, although certainly not 
complete, contains some of the features that would be present in an 
all-orders calculation in CHPT. In contrast, determinations of the 
$L_i^r$ from truncated NLO or NNLO CHPT expansions suffer from possibly 
large errors due to the neglect of higher-order contributions, and the 
poorly known values of the NNLO LECs.

Naturally, a further question to ask is whether the fitted value of 
$L_4^r$ could be modified by changing some of the features of the 
present description of the $J/\psi$ decays; Possible avenues of further 
investigation include the consideration of the $\eta\eta$ channel in 
the FSI which was neglected in this work, although it cannot be argued 
that this is necessarily a large effect as the description of the FSI 
is independent of the $L_i^r$ at this level of theoretical 
sophistication. It should also be noted that in the IAM framework of 
Ref.~\cite{IAM}, the $\pi\pi$ phase shift is given in terms of the 
$L_i^r$, and thus more flexibility of adjustment is possible. On the 
other hand, such an approach may actually turn out to be more 
constraining than the present one, since the pole position of the 
$f_0(980)$ resonance is then directly affected by the $L_i^r$.

Another essentially open question is whether a more relevant decay 
vertex than the one given in Eq.~(\ref{Lagr}) should be considered or 
not. In order to give a partial answer to this question, the pion 
scalar radius has been estimated for the present set of scalar~FFs. If 
the FFs were compensating heavily for some neglected momentum 
dependence in the $J/\psi \rightarrow \phi,\omega$ vertex, this might 
reveal itself as an unrealistic value for $\langle r^2\rangle_s^\pi$. 
The present analysis gives a value close to $\sim 0.65$~fm$^2$ which, 
as argued in Sect.~\ref{ff_res}, is in reasonable agreement with other 
recent theoretical predictions. It therefore appears that there is no 
compelling evidence as yet for a significant momentum dependence in the 
vertex described here by Eq.~(\ref{Lagr}).

An issue often raised is the different appearance of the dipion 
spectra in the $\omega\pi\pi$ and $\phi\pi\pi$ channels, which are 
shown in Fig.~\ref{fit_fig}. Within the present framework, this 
difference can be understood in very simple terms. According to the 
expectations from the OZI rule, the $\omega\pi\pi$ spectrum should be 
dominated by the nonstrange scalar~FF of the pion, whereas 
the $\phi\pi\pi$ process should essentially be driven by the strange 
scalar~FF alone. However, the latter one vanishes to 
LO in CHPT, which leads to a strong suppression of the low-energy tail 
of the dipion spectrum in $\phi\pi\pi$. On the other hand, the 
nonstrange~FF is nonzero to LO in CHPT, and thus the 
decay amplitude itself does not vanish for low dipion energies in the 
$\omega\pi\pi$ decay. The appearance of the experimental 
$\omega\pi\pi$ spectrum near the $\pi\pi$ threshold is then given 
by the squared modulus of the nonstrange scalar~FF and phase space.

In this context the OZI violation parameter $\lambda_\phi$ is 
instructive, since it determines whether any contribution from the 
nonstrange pion scalar~FF is present in the $\phi\pi\pi$ decay 
process. It turned out that $\lambda_\phi$ could be quite accurately 
determined from the present fits, which yielded a value of 
$\lambda_\phi = 0.13\pm 0.02$, which indicates that the nonstrange 
pion scalar~FF contributes significantly to the $\phi\pi\pi$ decay, and 
indeed this contribution appears as a distinct enhancement in the 
low-energy tail of the dipion distribution shown in Fig.~\ref{fit_fig}. 
Nevertheless, this parameter has decreased from the old 
value of around $\lambda_\phi \sim 0.17$ given in Ref.~\cite{UGM1}, due 
to the fact that the $S$-wave $\sigma$ contribution to the 
$\omega\pi\pi$ decay has turned out to be smaller than previously 
estimated. As discussed in Sect.~\ref{lphi_res}, the present value of 
$\lambda_\phi$ is consistent with that obtained in Ref.~\cite{Roca}, 
where the strong interaction in the $\omega\pi$ and $\phi\pi$ systems 
was explicitly taken into account. This was not done in the present 
work since a determination of the direct $S$-wave contribution via PWA 
is available~\cite{BES1,BES2,BES3}.

An extra bonus of the present work is that the dikaon distribution in 
the $\omega K^+K^-$ decay could be predicted close to the $K\bar K$ 
threshold, which is instructive since the present experimental data for 
that channel is difficult to interpret in terms of a 
PWA~\cite{BES3}. In the same way, the predicted shape of the 
$S$-wave distribution in the $\omega\pi^+\pi^-$ decay is slightly 
different from the one favored by the BES PWA. Finally, as the 
scalar~FFs of the pion and the kaon have been strongly constrained by 
the $J/\psi$ decays analyzed in this paper, the impact of these newly 
determined scalar~FFs can be studied for other processes like $B 
\rightarrow \sigma\pi$~\cite{Gardner:2001gc} or $B$ decays into three 
pseudoscalars, such as $B\rightarrow K^0_S\pi\pi$~\cite{Furman:2005xp}. 
Work along these lines is under way.

\begin{acknowledgments}
This research is part of the EU Integrated Infrastructure Initiative 
Hadron Physics Project under contract number RII3 - CT - 2004 - 506078 
and DFG (SBF/TR 16, ``Subnuclear Structure of Matter"). The authors 
thank David Bugg and Liaoyuan Dong for their invaluable assistance in 
making the BES data available. TL thanks Bachir Moussallam, Leonard 
Le\'sniak, Jos\'e Oller and Jos\'e Pel\'aez for email correspondence, 
and Johan Bijnens for assistance with the scalar FFs in CHPT, and 
Christoph Hanhart and Udit Raha for instructive discussions. TL is 
grateful to Bastian Kubis for sharing unpublished results on issues 
relevant to this paper. TL acknowledges a travel stipend from the 
Mikael Bj\"ornberg memorial foundation.
\end{acknowledgments}

\end{document}